\documentclass[journal,12pt]{IEEEtran}     
\usepackage{setspace}
\usepackage{cite}
\usepackage{amsmath,amssymb,amsfonts}
\usepackage{algorithmic}
\usepackage{graphicx}
\usepackage{textcomp}
\usepackage{xcolor}
\usepackage{mathtools}     
\usepackage{subcaption}   
\usepackage{hyperref}     
\usepackage{booktabs}  
\usepackage{placeins}
\usepackage{lineno}
\usepackage{multirow}
\newcommand*{\update}{\textcolor{black}} 
\newcommand*{\new}{\textcolor{black}} 
\newcommand*{\correct}{\textcolor{black}} 

\begin{document}

\title{Diffusion-based Surrogate Model for Time-varying Underwater Acoustic Channels}

\author{
    \IEEEauthorblockN{Kexin~Li\IEEEauthorrefmark{1}, Mandar~Chitre\IEEEauthorrefmark{1}\IEEEauthorrefmark{2}\\}
    \IEEEauthorblockA{\IEEEauthorrefmark{1}ARL, Tropical Marine Science Institute, National University of Singapore \\}
    \IEEEauthorblockA{\IEEEauthorrefmark{2}Department of Electrical and Computer Engineering, National University of Singapore \\}
}

\maketitle

\begin{abstract}
Accurate modeling of time-varying underwater acoustic channels is essential for the design, evaluation, and deployment of reliable underwater communication systems. Conventional physics models require detailed environmental knowledge, while stochastic replay methods are constrained by the limited diversity of measured channels and often fail to generalize to unseen scenarios, reducing their practical applicability. To address these challenges, we propose StableUASim, a pre-trained conditional latent diffusion surrogate model that captures the stochastic dynamics of underwater acoustic communication channels. Leveraging generative modeling, StableUASim produces diverse and statistically realistic channel realizations, while supporting conditional generation from specific measurement samples. Pre-training enables rapid adaptation to new environments using minimal additional data, and the autoencoder latent representation facilitates efficient channel analysis and compression. Experimental results demonstrate that StableUASim accurately reproduces key channel characteristics and communication performance, providing a scalable, data-efficient, and physically consistent surrogate model for both system design and machine learning-driven underwater applications. \new{In this work, we focus on sparse underwater channels, and expanding the diversity of the pre-training dataset is expected to further improve generalization to a broader range of underwater propagation environments.}
\end{abstract}


\section{Introduction}
Underwater acoustic communication~(UAC) is critical for applications such as ocean exploration, environmental monitoring, defence, and operation of autonomous underwater vehicles~\cite{lurton2002introduction}. Reliable UAC remains challenging because underwater channels exhibit severe multipath propagation, frequency-selective fading, limited bandwidth, and strong Doppler effects. These characteristics depend on environmental factors such as bathymetry, surface conditions, and sound speed profiles, resulting in dynamic, non-stationary, and time-varying impulse responses~(TVIRs)~\cite{stojanovic2009underwater}. A TVIR describes the temporal evolution of a channel's multipath structure, capturing both short-term fading dynamics and long-term statistical variability. Therefore, TVIRs provide a faithful representation of UAC channels. Accurate modeling of TVIRs is useful in the design, evaluation, and deployment of robust underwater communication systems. Furthermore, with the increasing adoption of machine learning~(ML) in underwater acoustics, realistic channel models are necessary for generating training datasets for tasks such as localization~\cite{yuan2025research} and adaptive transmission, both of which critically depend on capturing the variability and dynamics of real-world channels.

Traditional approaches to UAC modeling can be broadly categorized into physics methods~\cite{etter2018underwater} and replay methods~\cite{otnes2013validation}. Physics methods, such as ray-tracing~\cite{gul2017underwater} and normal-mode methods~\cite{westwood1996normal}, provide approximate solutions to the acoustic wave equation under specific assumptions regarding frequency, propagation range, or boundary conditions. Ray-tracing treats acoustic propagation as geometric rays, which is accurate at high frequencies, while normal-mode methods decompose the wavefield into discrete modes, making them effective in low-frequency scenarios. Although these models are physically interpretable and capable of producing high-fidelity outputs when provided with accurate inputs, they depend on detailed environmental parameters~(e.g., bathymetry, sound speed profiles, and seabed properties) and can be computationally intensive. In practice, the difficulty of acquiring precise oceanographic data limits their applicability for channel modeling. Furthermore, these models do not model the time-variability of the channels directly, and require some statistical models of time variability to be adopted to generate TVIRs. By contrast, replay methods reconstruct channel dynamics by reusing measured TVIRs to produce new realizations that preserve key statistical characteristics, including delay spread, Doppler, and path-dependent fading. Common approaches include direct replay~\cite{otnes2013validation}, which deterministically reproduces subsequent channel responses from measured probes, and stochastic replay~\cite{socheleau2015stochastic,socheleau2015parametric}, which introduces controlled randomness to generate statistically consistent variations beyond the original measurements. Although replay methods achieve high physical fidelity without requiring detailed environmental knowledge, they inherently lack generalization: they cannot extrapolate to unobserved environments, adapt to new deployment scenarios, or generate channel realizations beyond the diversity captured in the available measurements. Consequently, repeated and costly measurements are often required for each new environment.

These limitations have motivated the development of generative  ML surrogate models for UAC channels. By learning the underlying statistical distributions of channel responses from measured data, generative models can produce an unlimited number of realistic realizations that capture channel dynamics, going beyond the simple reuse of existing measurements and potentially generalizing to previously unobserved conditions. Prior studies have employed generative adversarial networks~(GANs)~\cite{10608448} and diffusion models~\cite{10847985} to construct channel surrogate models that capture complex temporal correlations in TVIRs. Despite these advances, existing generative models are typically trained on specific datasets, which limits their ability to generalize to unseen environments and constrains conditional controllability. Developing models capable of robust, physically consistent, and controllable channel generation across diverse environments remains a critical challenge for practical UAC modeling.

To address these challenges, we propose a pre-trained conditional latent diffusion surrogate model for UAC channels. The proposed approach combines the flexibility of generative modeling with enhanced data efficiency and offers several key advantages:
\begin{itemize}
    \item Stochastic modeling: captures multipath propagation, fading, and Doppler dynamics, enabling diverse channel realizations.
    \item Data efficiency: pre-training on diverse environments allows rapid adaptation to a new target environment using only a small number of measurements.
    \item Conditional controllability: supports channel generation conditioned on specific measurement samples.
    \item Statistical realism: produces channels that accurately reproduce key channel metrics and end-to-end communication performance, validated through experiments.
    \item Latent representation learning: encodes TVIRs into a compact latent space via an autoencoder, facilitating efficient analysis and providing a useful by-product for channel compression.
\end{itemize}
Collectively, these features yield a scalable, data-efficient, and generalizable surrogate channel model for UAC, enabling robust system design, accurate performance evaluation, and support for \correct{ML} applications in the underwater domain. \update{Accordingly, in this work, we focus on sparse underwater channels, which are representative of typical multipath conditions in many underwater acoustic scenarios and consistent with the channel characteristics used during pre-training.}

\section{Problem and Solution Formulation}
\label{sec:problem_formulation}

Modeling UAC channels requires capturing the stochastic and time-varying nature of TVIRs, which exhibit strong temporal correlations and variability that are critical to communication system performance. Conventional channel modeling approaches often rely on detailed oceanographic parameters and extensive real-world measurements, which are costly, environment-specific, difficult to scale, and may fail to generalize to unseen environments. To address these challenges, we formulate UAC channel modeling as a conditional generative problem: given an observed TVIR $\mathbf{X}_c$ from a target environment, the objective is to generate an arbitrary number of realistic TVIRs that preserve both temporal dynamics and statistical characteristics relevant to communication performance. A key requirement is data efficiency: the model must adapt to new environments using only a small number of measurements. Our solution is a conditional diffusion surrogate model pre-trained on a large and diverse collection of simulated channels. Pre-training equips the model with a broad understanding of the stochastic, time-varying structure of UAC channels, providing prior knowledge that enables rapid adaptation to new environments with minimal fine-tuning. Once adapted, the surrogate can generate an arbitrary number of realistic channel realizations, suitable for evaluating and benchmarking communication system performance in the target environment.

Formally, let $\mathbf{X} \in \mathbb{C}^{T \times D}$ denote a TVIR instance, where $T$ is the number of time snapshots and $D$ is the number of delay taps per snapshot. Each TVIR captures the temporal evolution of a UAC channel over a short duration. Our surrogate model is formulated as a generative model, which aims to learn the underlying distribution of channel responses and generate realistic TVIR realizations conditioned on observed measurements from a target environment. Specifically, given an observed TVIR $\mathbf{X}_c$ from the target environment, the model defines a conditional distribution over TVIRs:
\begin{equation}
p_{\boldsymbol{\theta}}(\mathbf{X} \mid \mathbf{X}_c),
\end{equation}
where $\boldsymbol{\theta}$ denotes the model parameters. Channel realizations are obtained by sampling from this distribution:
\begin{equation}
\hat{\mathbf{X}} \sim p_{\boldsymbol{\theta}}(\mathbf{X} \mid \mathbf{X}_c),
\end{equation}
with $\hat{\mathbf{X}}$ representing a generated TVIR that reflects the statistical properties of the target environment.

To enable rapid adaptation to new target environments with limited measurements, the model is first pre-trained on a diverse set of TVIRs $\{\mathbf{X}_i\}_{i=1}^N$ sampled from simulated environments to capture the general stochastic structure of underwater acoustic channels. The optimal pre-trained model parameters are denoted as $\boldsymbol{\theta}^*_{\text{pre-trained}}$:
\begin{equation}
\boldsymbol{\theta}^*_{\text{pre-trained}} = \arg\min_{\boldsymbol{\theta}} \mathcal{L}(\boldsymbol{\theta}; \{\mathbf{X}_i\}_{i=1}^N),
\end{equation}
where $\mathcal{L}$ denotes the chosen training loss function. The pre-trained model can then be employed in two modes:

\begin{enumerate}
    \item Zero-shot generation: generate channel realizations conditioned on a measurement $\mathbf{X}_c$ from the target environment without updating the pre-trained parameters:
    \begin{equation}
    \hat{\mathbf{X}} \sim p_{\boldsymbol{\theta}^*_{\text{pre-trained}}}(\mathbf{X} \mid \mathbf{X}_c).
    \end{equation}

    \item Fine-tuning: adapt the model parameters $\boldsymbol{\theta}$ to the target environment using a limited set of measured TVIRs $\{\mathbf{X}_j^{\text{meas}}\}_{j=1}^M$:
    \begin{equation}
    \boldsymbol{\theta}^*_{\text{fine-tuned}} = \arg\min_{\boldsymbol{\theta}} 
    \mathcal{L}(\boldsymbol{\theta}; \{\mathbf{X}_j^{\text{meas}}\}_{j=1}^M),
    \end{equation}
    where $\boldsymbol{\theta}$ is initialized from $\boldsymbol{\theta}^*_{\text{pre-trained}}$. Channel generation is then performed conditioned on $\mathbf{X}_c$:
    \begin{equation}
    \hat{\mathbf{X}} \sim p_{\boldsymbol{\theta}^*_{\text{fine-tuned}}}(\mathbf{X} \mid \mathbf{X}_c).
    \end{equation}
\end{enumerate}

TVIRs are typically high-dimensional complex-valued matrices. Rather than training the conditional generative model directly on raw data (in our case, complex matrices representing the TVIRs), a common strategy is to compress data into a compact latent space and perform all training in this latent space\update{~\cite{doi:10.1126/science.1127647,dantas2024comprehensive,10.1109/TPAMI.2013.50}}. This typically improves training efficiency and reduces model complexity. Specifically, an autoencoder with encoder $\mathcal{E}(\cdot)$ and decoder $\mathcal{D}(\cdot)$ is trained to map TVIRs to a low-dimensional latent representation:
\begin{equation}
\mathbf{z} = \mathcal{E}(\mathbf{X}) \in \mathbb{R}^L, \quad L \ll T \times D,
\end{equation}
and to reconstruct the TVIR via
\begin{equation}
\hat{\mathbf{X}} = \mathcal{D}(\mathbf{z}).
\end{equation}
In the context of our problem, the conditional generative model is trained to capture the distribution of the latent variable $\mathbf{z}$ conditioned on the latent embedding of the observed TVIR, $\mathbf{z}_c = \mathcal{E}(\mathbf{X}_c)$:
\begin{equation}
    \hat{\mathbf{z}} \sim p_{\boldsymbol{\theta}}(\mathbf{z} \mid \mathbf{z}_c),
\end{equation}
and decoding the generated latent samples yields realistic TVIRs:
\begin{equation}
\hat{\mathbf{X}} = \mathcal{D}(\hat{\mathbf{z}}).
\end{equation}

This conditional latent generative framework, tailored to our problem, provides a flexible, data-driven, and environment-aware surrogate model for UAC channels, enabling efficient generation of realistic TVIRs for evaluation and analysis in target environments.

\section{Model Architecture}
\subsection{Overview}

Generative models have shown great promise for learning complex, high-dimensional data distributions. Among common paradigms, GANs can produce sharp and realistic samples, making them effective for high-fidelity synthesis~\cite{goodfellow2020generative,creswell2018generative, wang2017generative}. However, GANs are prone to training instability and mode collapse, which can limit sample diversity. \correct{Variational autoencoders}~(VAEs)~\correct{\cite{kingma2019introduction}}, on the other hand, offer stable training and a probabilistic, interpretable latent space, but standard VAEs often produce slightly blurred outputs and may fail to capture fine-grained temporal or spatial structures~\cite{kingma2019introduction,pinheiro2021variational}. Diffusion models address these limitations by generating data through an iterative denoising process, combining stable training, high-fidelity generation, and the ability to capture complex, multimodal distributions~\cite{ho2020denoising,cao2024survey}. Although computationally more expensive due to iterative sampling, they are particularly effective for high-dimensional, temporally correlated, and non-stationary data such as TVIRs, overcoming the trade-offs of both GANs and VAEs.

\begin{figure}
	\centering
 	\includegraphics[width=0.5\textwidth]{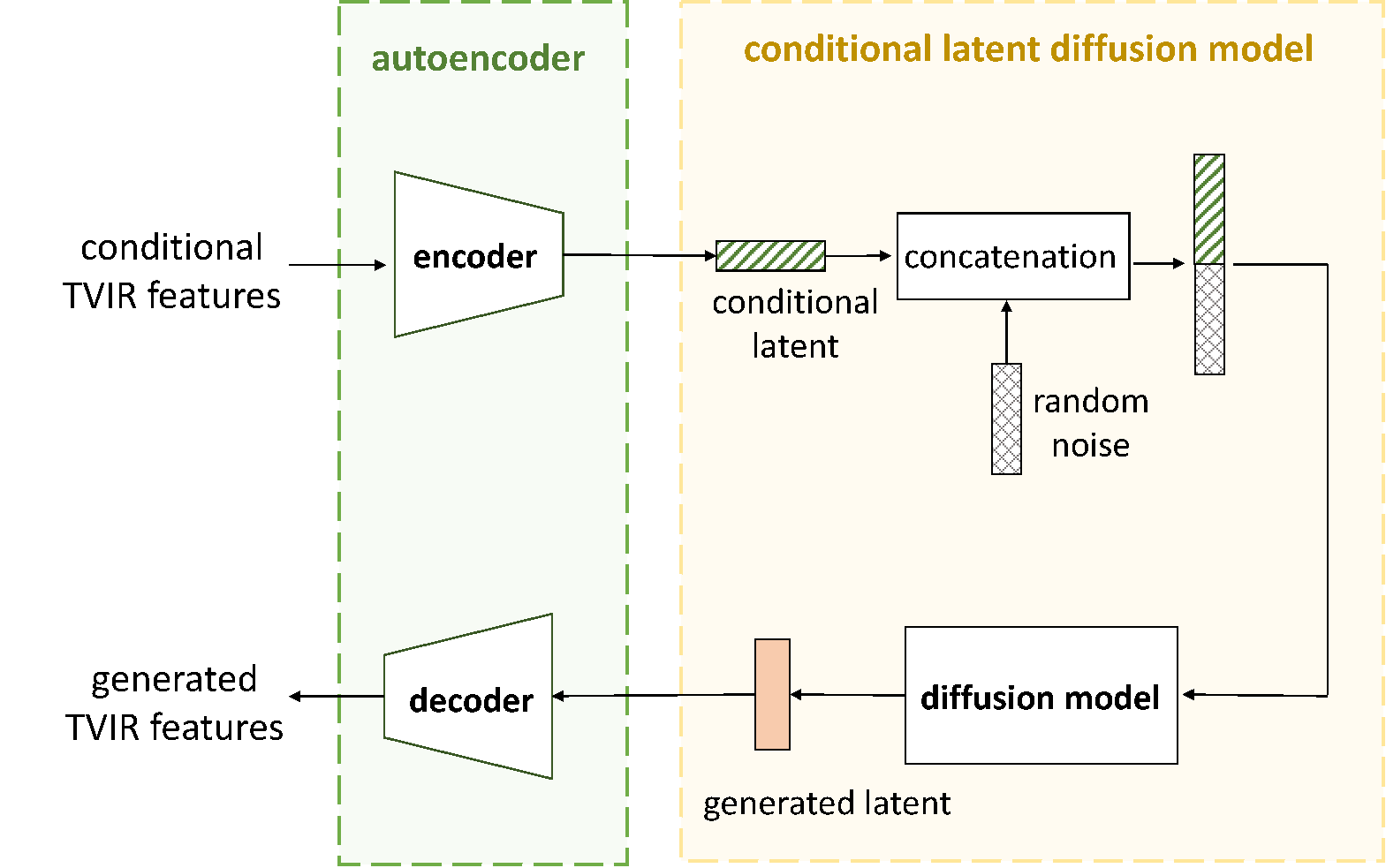}
	\caption{Architecture of the proposed StableUASim model for TVIR generation.}
	\label{fig:system_arc}
\end{figure}

Building on the general formulation presented in Section~\ref{sec:problem_formulation}, we leverage the generative power of diffusion models to propose a conditional latent diffusion approach for surrogate channel modeling, capable of generating realistic TVIRs. We term this model \emph{StableUASim}~(Stable diffusion Underwater Acoustic Simulator). Fig.~\ref{fig:system_arc} illustrates the proposed StableUASim model structure, which combines an encoder-decoder mapping with a latent-space conditional diffusion module. We adopt TVIRs as model input, as they retain temporal correlations and the inherent sparsity of multipath components, which are essential for capturing non-stationary channel dynamics. While prior work~\cite{10847985} employed Fourier-transformed channel responses, there is no clear evidence that frequency-domain representations provide advantages over their time-domain counterparts.

To demonstrate the approach in a realistic setting, we configure the input TVIRs using measurements representative of Singapore waters, where environmental dynamics evolve slowly and a maximum delay spread of approximately 20~ms captures the key channel characteristics. Each TVIR is sampled at 20~Hz along the time axis and 12~kHz along the delay axis, spanning 1~s in time and approximately 20.83~ms in delay, resulting in $20 \times 250$ complex-valued elements per TVIR. Each temporal slice of a TVIR corresponds to a channel impulse response~(CIR). The choice of a 1~s window ensures that temporal channel evolution is adequately captured while keeping the input dimensionality manageable for model training. This duration is also well-aligned with our target applications \update{of adaptive modulation in underwater acoustic modems}, where transmitted communication packets are typically about 1~s long. Although this configuration is used for training and evaluation in this paper, the framework is general and can be adapted to TVIRs of different durations or sampling rates by appropriately adjusting the model architecture.

\subsection{Autoencoder }

The autoencoder is designed to compress TVIRs while preserving essential temporal and structural features. TVIRs consist of sequences of CIRs that exhibit strong correlation across consecutive snapshots. Moreover, UAC channels are typically sparse, with only a few significant multipath components present at each time instant. These properties of sparsity and temporal correlation facilitate effective compression. To exploit this, both the encoder and decoder employ Long Short-Term Memory~(LSTM) networks, which are well-suited for modeling sequential data with long-term dependencies~\cite{greff2016lstm}.

In our implementation, CIR snapshots are first downsampled from the recorded 12~kHz signal to 20~Hz along the time dimension before encoding to reduce computational cost, and subsequently upsampled for evaluation of communication systems. The autoencoder compresses each $20 \times 250$ complex-valued TVIR into a 128-dimensional real-valued latent representation, corresponding to a compression ratio of approximately\footnote{Calculated as $20 \times 250 \times 2 / 128 \approx 78$.} 78$\times$. When accounting for temporal downsampling, the effective overall compression ratio\footnote{Calculated as $(12,000 / 20) \times 78 \approx 46,800$, where 12,000 is the original sampling rate in Hz, 20 is the downsampled sampling rate, and 78 is the latent compression factor.} reaches 46,800$\times$, while still maintaining good fidelity in terms of the ability to predict communication performance.

Although complex-valued neural networks can naturally model the relationships between amplitude and phase of complex inputs, they are less explored~\cite{lee2022complex}, harder to train, and not natively supported in popular frameworks such as PyTorch or TensorFlow. To address this, we convert complex-valued TVIRs $\mathbf{X}$ into a real-valued form suitable for standard LSTM networks. Recall Euler's formula, which represents a complex number as:
\begin{equation}
e^{i\phi} = \cos\phi + i \sin\phi,
\end{equation}
indicating that phase information can be expressed using bounded and continuous sine and cosine functions. Following this principle, each CIR $\mathbf{x}_i$ in a TVIR $\mathbf{X}$ is mapped to a real-valued feature vector $\mathbf{h}_i$ by concatenating its amplitude $\mathbf{A}_i$ with the phase-related terms $\sin\boldsymbol{\phi}_i$ and $\cos\boldsymbol{\phi}_i$:
\begin{equation}
\mathbf{h}_i = [\mathbf{A}_i, \sin\boldsymbol{\phi}_i, \cos\boldsymbol{\phi}_i],
\label{eqn:h}
\end{equation}
where $\mathbf{A}_i$ and $\boldsymbol{\phi}_i$ denote the amplitude and phase of the $i^\text{th}$ CIR, respectively. This representation jointly preserves amplitude and phase while avoiding discontinuities associated with direct phase unwrapping. \update{We also evaluated an alternative representation by concatenating the real and imaginary components. However, it produced less consistent feature scaling and inferior performance, supporting the use of the magnitude-phase representation.}
For each CIR $\mathbf{x}_i$ with 250 delay taps, the resulting feature vector $\mathbf{h}_i$ has dimension 750 and is normalized to a relaxed\footnote{We normalize each TVIR by the element with the maximum amplitude in its first CIR. This relaxed normalization allows amplitudes in subsequent CIRs to slightly exceed 1, preserving temporal variations across the TVIR while keeping values in a range that facilitates model learning.} range of 0-1.

Each TVIR $\mathbf{X}$ is represented as a sequence of these feature vectors:
\begin{equation}
\mathbf{H} = [\mathbf{h}_1, \mathbf{h}_2, \dots, \mathbf{h}_{20}],
\end{equation}
which is encoded into a latent vector via the encoder network:
\begin{equation}
\mathbf{z} = \mathcal{E}_{\boldsymbol{\theta}_\text{E}}(\mathbf{H}),
\end{equation}
and decoded back to the feature space by the decoder network:
\begin{equation}
\hat{\mathbf{H}} = \mathcal{D}_{\boldsymbol{\theta}_\text{D}}(\mathbf{z}),
\end{equation}
where $\hat{\mathbf{H}} = [\hat{\mathbf{h}}_1, \hat{\mathbf{h}}_2, \dots, \hat{\mathbf{h}}_{20}]$ consists of the reconstructed CIR feature vectors, and $\mathcal{E}_{\boldsymbol{\theta}_\text{E}}(\cdot)$ and $\mathcal{D}_{\boldsymbol{\theta}_\text{D}}(\cdot)$ are parameterized by weights $\boldsymbol{\theta}_\text{E}$ and $\boldsymbol{\theta}_\text{D}$.

Each decoded feature vector $\hat{\mathbf{h}}_i$ is split into amplitude and phase components:
\begin{equation}
\mathbf{\hat{h}}_i = [\hat{\mathbf{A}}_i, \sin\boldsymbol{\hat{\phi}}_i, \cos\boldsymbol{\hat{\phi}}_i],
\end{equation}
and the original phase is recovered as:
\begin{equation}
\boldsymbol{\hat{\phi}}_i = \tan^{-1}\Big(\frac{\sin\boldsymbol{\hat{\phi}}_i}{\cos\boldsymbol{\hat{\phi}}_i}\Big).
\label{eqn:hhat}
\end{equation}

Finally, the reconstructed CIR in complex-valued form is obtained from each decoded feature vector:
\begin{equation}
\hat{\mathbf{x}}_i = \hat{\mathbf{A}}_i e^{i \, \boldsymbol{\hat{\phi}}_i},
\end{equation}
and the full reconstructed TVIR is:
\begin{equation}
\mathbf{\hat{X}} = [\hat{\mathbf{x}}_1, \hat{\mathbf{x}}_2, \dots, \hat{\mathbf{x}}_{20}].
\end{equation}

\begin{figure*}
	\centering
	\begin{subfigure}{0.75\textwidth}
 		\includegraphics[width=\textwidth]{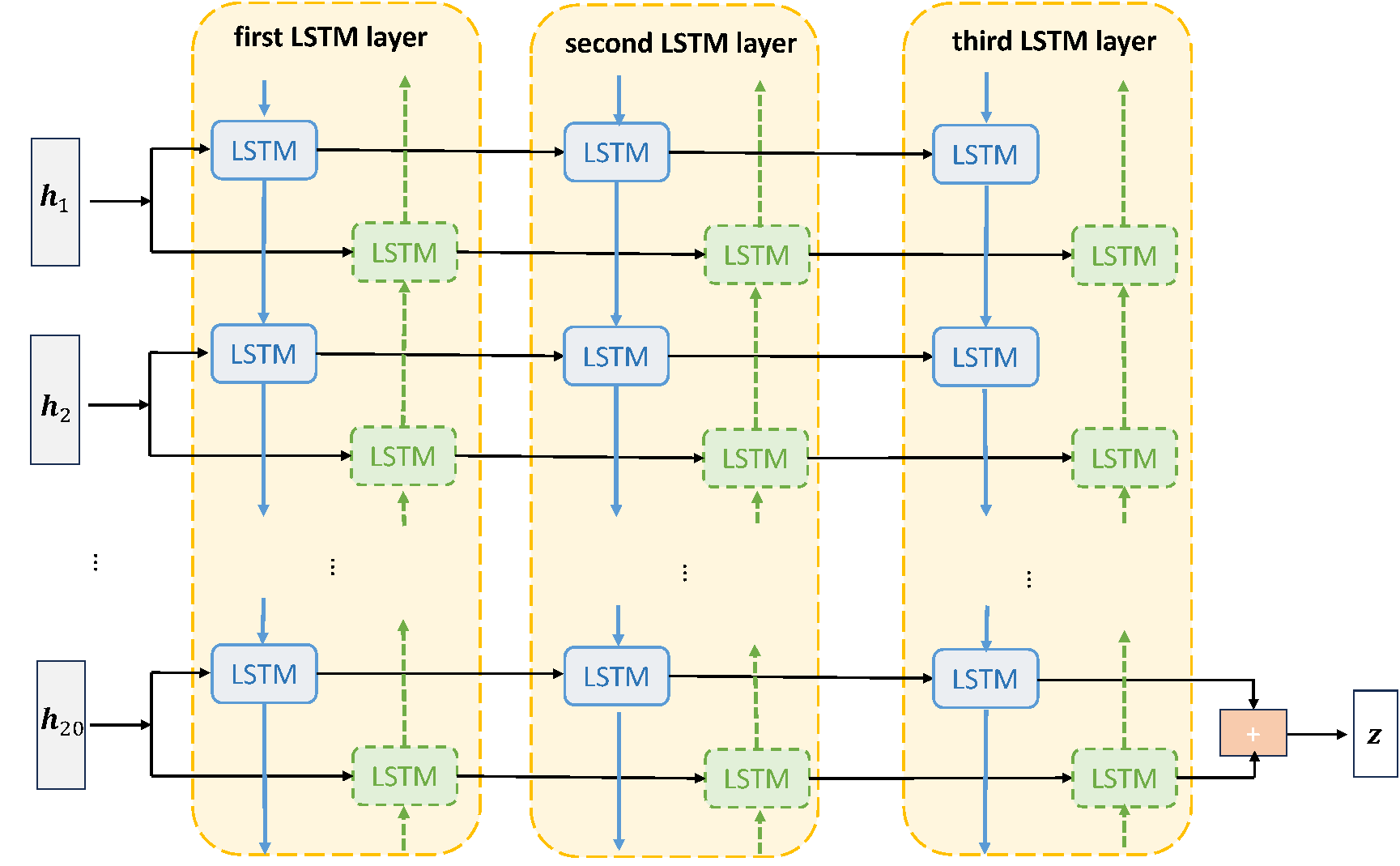}
    	\caption{Bi-LSTM TVIR encoder.}
    \hfill
	\end{subfigure}
	\begin{subfigure}{0.75\textwidth}
\includegraphics[width=\textwidth]{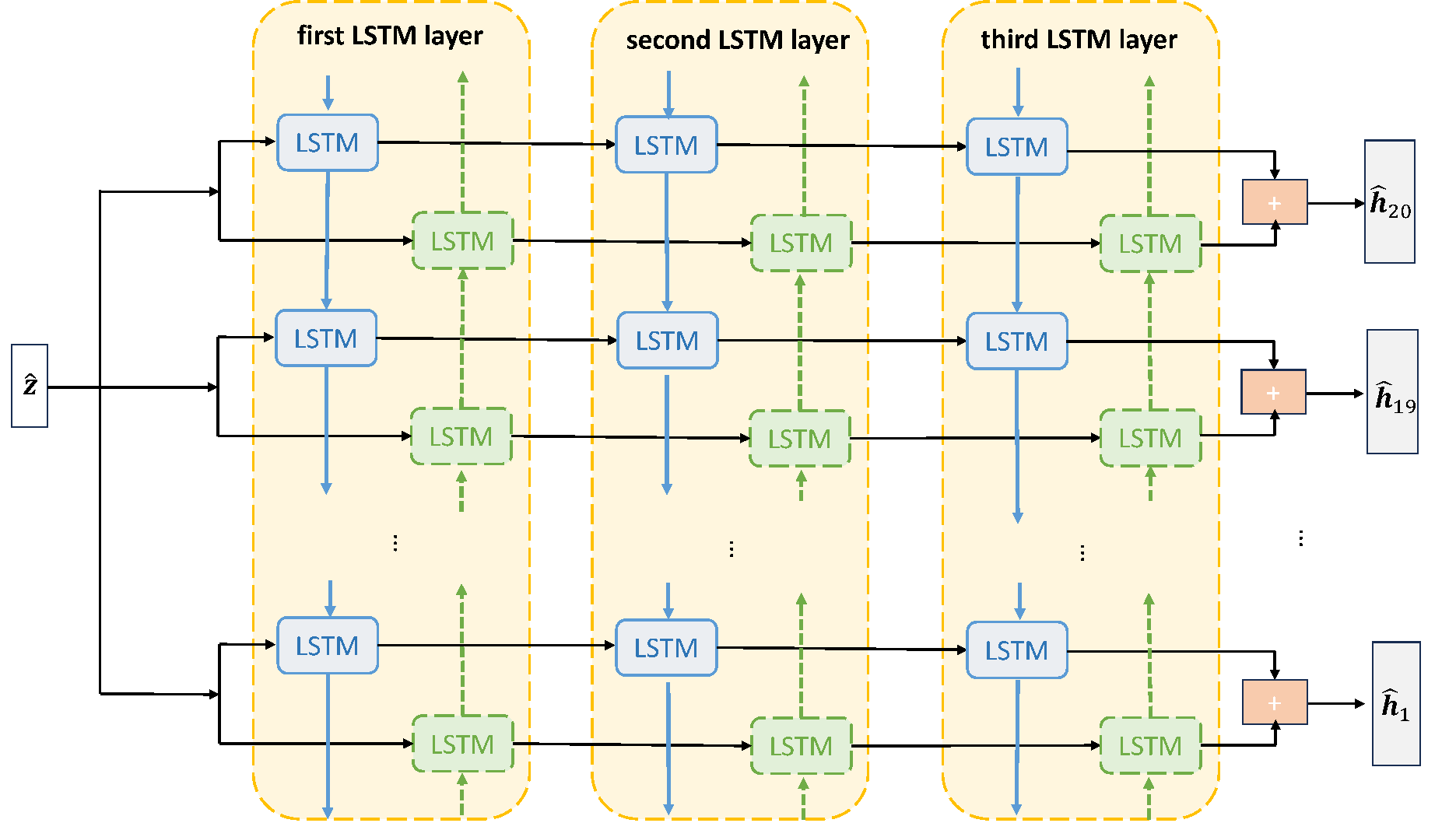}
    	\caption{Bi-LSTM TVIR decoder.}
	\end{subfigure}
	\caption{Bi-LSTM autoencoder architecture used in StableUASim. Each CIR feature $\mathbf{h}_i$ is encoded into a 128-dimensional latent vector $\mathbf{z}$ and reconstructed as $\hat{\mathbf{h}}_i$. During autoencoder training, $\hat{\mathbf{z}} = \mathbf{z}$, while in the diffusion model, $\hat{\mathbf{z}}$ is generated conditionally. The Bi-LSTM processes the sequence bidirectionally: forward (solid blue line, first to last element) and backward (dashed green line, last to first element). Hidden states from both directions are combined at each step. The three stacked LSTM layers share weights across time steps, capturing long-range temporal correlations in TVIR sequences.}
	\label{fig:LSTM_AE_bi}
\end{figure*}

\begin{figure*}
\centering
\includegraphics[width=1\textwidth]{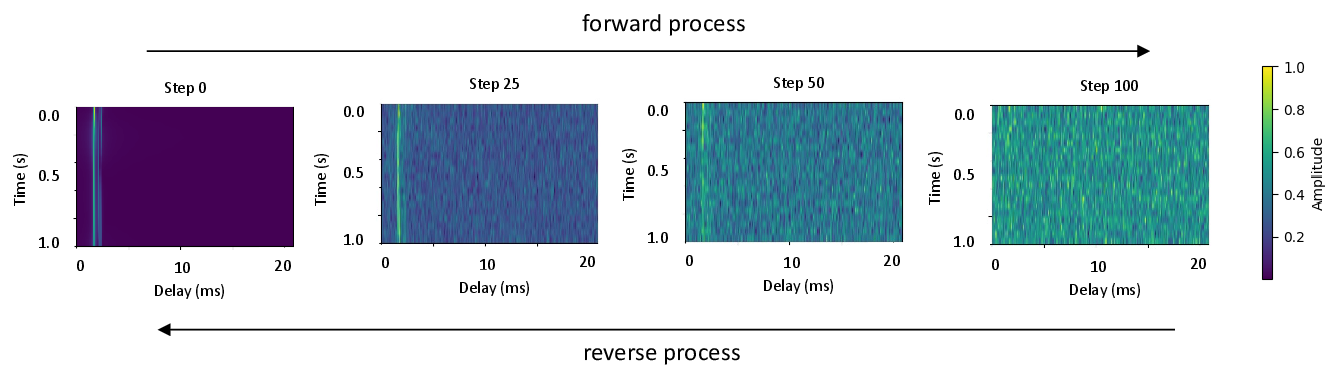}
\caption{Example illustrating the forward and reverse processes of a diffusion model applied to a TVIR.}
\label{fig:diff}
\end{figure*}

To effectively encode and decode the sequential TVIR features described above, we employ a three-layer stacked bidirectional LSTM (Bi-LSTM) for both the encoder and decoder. The encoder maps the $20 \times 750$ feature sequence of each TVIR into a 128-dimensional latent vector by combining hidden states from both directions. The decoder mirrors this structure, reconstructing the original $20 \times 750$ feature sequence by combining forward and backward outputs. All model states are initialized to zero. This bidirectional design enables the model to leverage both past and future context, capturing long-range temporal correlations critical for modeling non-stationary underwater channels. The overall architecture is illustrated in Fig.~\ref{fig:LSTM_AE_bi}. \update{This three-layer Bi-LSTM balances model capacity and computational efficiency for TVIR representation learning. The required capacity is mainly influenced by channel characteristics such as delay spread and Doppler spread, which govern the structural complexity and temporal dynamics of the input sequence. Larger delay spreads increase multipath complexity, while stronger Doppler effects introduce faster temporal variations, potentially requiring higher hidden dimensions or deeper temporal modeling.}

The autoencoder is trained with a composite loss that jointly accounts for amplitude and phase reconstruction, emphasizing significant multipath components while downweighting near-zero amplitudes. The goal is to learn the optimal encoder and decoder parameters, $\boldsymbol{\theta}^{\ast}_E$ and $\boldsymbol{\theta}^{\ast}_D$, by solving:
\begin{equation}
\boldsymbol{\theta}^{\ast}_E, \boldsymbol{\theta}^{\ast}_D = \operatorname*{arg\,min}_{\boldsymbol{\theta}_E, \boldsymbol{\theta}_D}  \mathcal{L}_{\text{AE}}(\boldsymbol{\theta}_E, \boldsymbol{\theta}_D; \mathbf{H}),
\end{equation}
where the per-sample loss is defined as:
\begin{equation}
\mathcal{L}_{\text{AE}}(\boldsymbol{\theta}_E, \boldsymbol{\theta}_D;\mathbf{H}) 
= \mathcal{L}_{\text{amp}}(\mathbf{H},\mathbf{\hat{H}}) 
+ \eta \, \mathcal{L}_{\text{phase}}(\mathbf{H}, \mathbf{\hat{H}}),
\end{equation}
with:
\begin{subequations}
\begin{align}
\mathbf{\hat{H}} &= \mathcal{D}_{\boldsymbol{\theta}_\text{D}}(\mathcal{E}_{\boldsymbol{\theta}_\text{E}}(\mathbf{H})),\\
\mathcal{L}_{\text{amp}}(\mathbf{H}, \mathbf{\hat{H}}) 
&= \sum_{i=1}^{20}\sum_{j=1}^{250} 
\big| A_{i,j} - \hat{A}_{i,j} \big|^2, \\
\mathcal{L}_{\text{phase}}(\mathbf{H},\mathbf{\hat{H}}) 
&= \sum_{i=1}^{20}\sum_{j=1}^{250} \Big( 
\big| A_{i,j}\sin(\phi_{i,j}) - \hat{A}_{i,j}\sin(\hat{\phi}_{i,j}) \big|^2 \notag \\
&\quad + \big| A_{i,j}\cos(\phi_{i,j}) - \hat{A}_{i,j}\cos(\hat{\phi}_{i,j}) \big|^2 \Big),
\end{align}
\end{subequations}
where $\eta$ is the \new{amplitude-phase} weighting factor (set to $\eta=1$ during training), and $A_{i,j}$ and $\phi_{i,j}$ are the ground truth amplitude and phase of the $j^\text{th}$ tap in the $i^\text{th}$ CIR. The loss is computed per sample and aggregated over the batch. \update{It comprises a magnitude error term, which provides explicit supervision for accurate amplitude reconstruction, and a magnitude-weighted phase error term, which emphasizes phase errors of high-amplitude taps by using magnitude as a weighting factor. Including both terms enables effective learning of both amplitude and phase, while improving training stability and convergence through direct magnitude supervision.}

\subsection{Conditional Latent Diffusion Model}

Diffusion models are a state-of-the-art generative framework for synthesizing high-fidelity and diverse samples~\cite{ho2020denoising,cao2024survey,yang2023diffusion}. These models consist of two complementary processes: a \emph{forward process} that gradually adds Gaussian noise to a sample until it becomes indistinguishable from isotropic noise, and a \emph{reverse process}, parameterized by a neural network, which iteratively denoises the corrupted sample to reconstruct the original data~\cite{rombach2022high}. Fig.~\ref{fig:diff} illustrates these processes applied to a TVIR.

Formally, the forward diffusion process perturbs a latent variable $\mathbf{z}_{t-1}$ according to:
\begin{equation}
q(\mathbf{z}_t | \mathbf{z}_{t-1}) = \mathcal{N}(\mathbf{z}_t; \sqrt{1-\beta_t}\mathbf{z}_{t-1}, \beta_t \mathbf{I}),
\end{equation}
which can be reparameterized as:
\begin{equation}
\mathbf{z}_t = \sqrt{1-\beta_t}\mathbf{z}_{t-1} + \sqrt{\beta_t}\boldsymbol{\epsilon}_{t-1},
\label{eqn:zt}
\end{equation}
where $\boldsymbol{\epsilon}_{t-1} \sim \mathcal{N}(0, \mathbf{I})$, $t = 1, \dots, T$ indexes the diffusion time steps, and $\mathbf{I}$ denotes the identity matrix. 
The sequence $\{\beta_t\}_{t=1}^T$ defines the noise schedule~\cite{chen2023importance}, which controls the magnitude of Gaussian perturbations added at each step.

Recursively, this yields the closed-form expression:
\begin{equation}
\mathbf{z}_t = \sqrt{\bar{\alpha}_t} \mathbf{z}_0 + \sqrt{1-\bar{\alpha}_t} \boldsymbol{\epsilon},
\end{equation}
where 
\begin{equation}
\bar{\alpha}_t = \prod_{s=1}^t (1-\beta_s).
\end{equation}

The core of the reverse diffusion process is the denoising network, which iteratively removes noise to recover the clean latent representation. In our framework, this network $\kappa(\cdot)$ predicts the noise component at each time step $t$, taking as input the noisy latent $\mathbf{z}_t$, a 32-dimensional positional embedding of $t$~\cite{vaswani2017attention,ho2020denoising}, and a conditional latent $\mathbf{z}_c$ derived from an observed TVIR. This design allows the model to generate realistic, condition-guided channel responses.

\begin{figure}[!ht]
    \centering
    \begin{subfigure}{\columnwidth}
        \centering
        \includegraphics[width=\columnwidth]{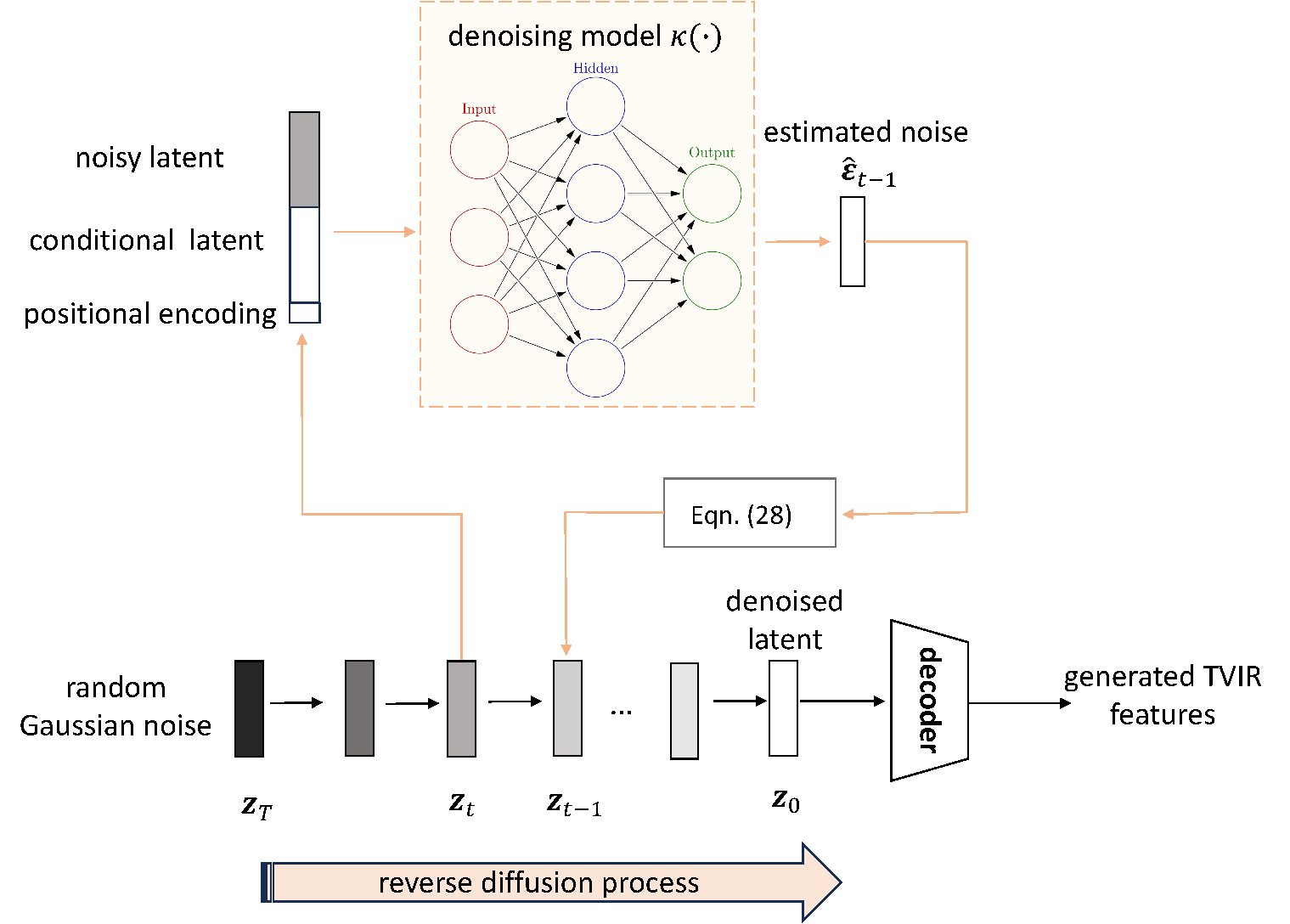}
        \caption{Reverse process of the diffusion model.}
        \label{fig:diff_reverse}
    \end{subfigure}

    \vspace{0.5cm} 
    \begin{subfigure}{\columnwidth}
        \centering
        \includegraphics[width=\columnwidth]{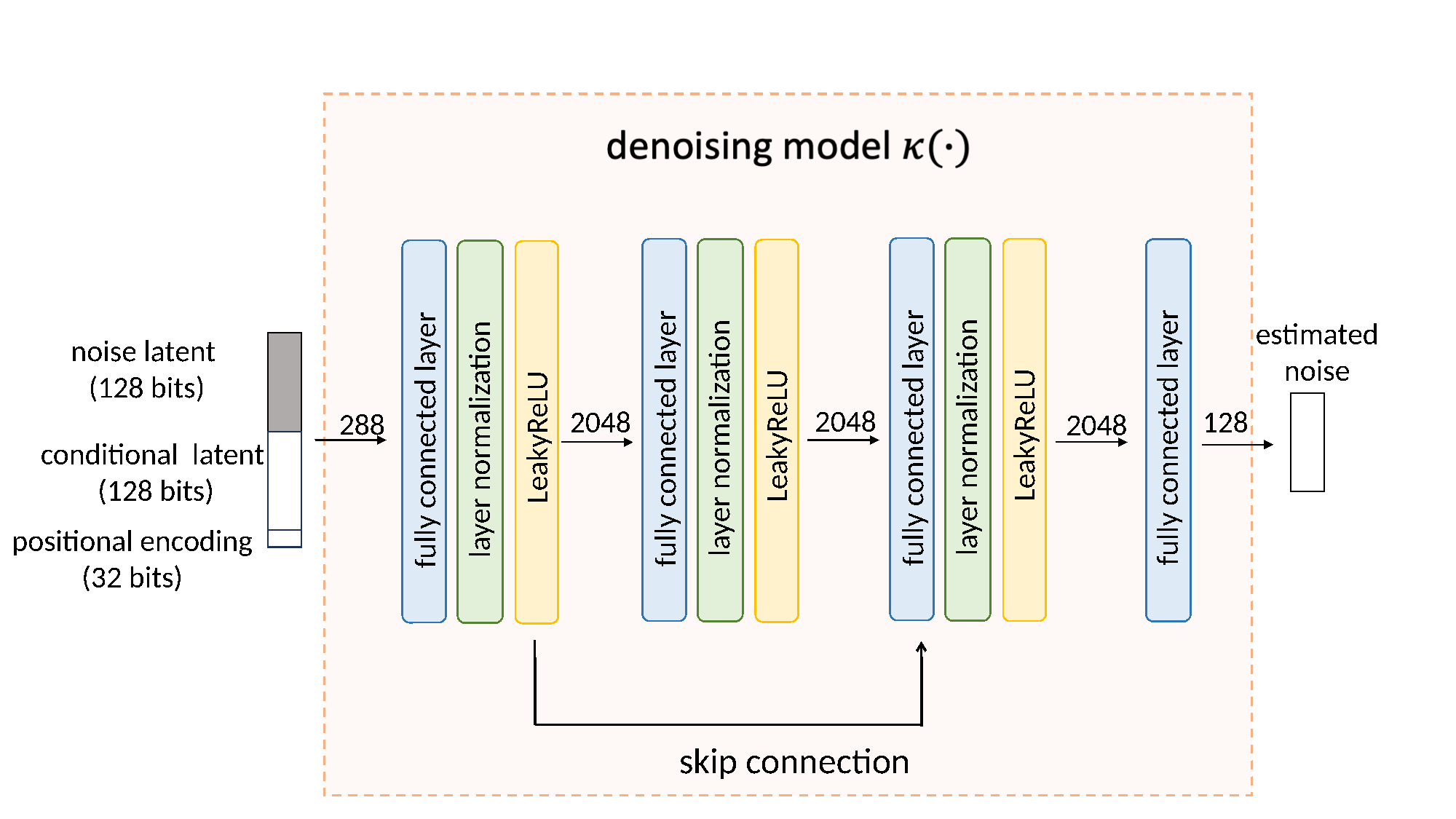}
        \caption{Denosing network architecture, showing feature dimensions at each stage.}
        \label{fig:diff_arch}
    \end{subfigure}
    \caption{Overview of the proposed conditional latent diffusion model.}
    \label{fig:diff_overview}
\end{figure}

As illustrated in Fig.~\ref{fig:diff_overview}, $\kappa(\cdot)$ is implemented as a fully connected network with residual connections to preserve latent features and mitigate vanishing gradients. It consists of three hidden layers of 2048 units each, equipped with layer normalization and LeakyReLU activations, followed by a residual bridge that facilitates gradient flow. The network outputs a 128-dimensional predicted noise vector. Starting from a random Gaussian latent $\mathbf{z}_T$, the reverse diffusion process iteratively applies this denoising network to obtain the clean latent $\mathbf{z}_0$, which is finally decoded into the TVIR domain via the decoder $\mathcal{D}(\cdot)$. \update{The specific architecture and hyperparameters of this latent diffusion model were determined through empirical evaluation, balancing reconstruction fidelity and computational efficiency. Designed to operate on the latent representations produced by the LSTM autoencoder, it leverages the compressed yet informative features to accurately reconstruct the temporal and delay structures of the TVIR.}

Formally, at each step $t$, the reverse process predicts the noise term using the neural network $\kappa(\cdot)$, parameterized by weights $\boldsymbol{\theta}_\text{diff}$:
\begin{equation}
\hat{\boldsymbol{\epsilon}}_{t-1} = \kappa_{\boldsymbol{\theta}_\text{diff}}(\mathbf{z}_t, \mathbf{z}_c, t),
\end{equation}
and denoises the latent one step further as:
\begin{equation}
\mathbf{z}_{t-1} = \frac{1}{\sqrt{\alpha_t}}\Big(\mathbf{z}_t - \frac{\beta_t}{\sqrt{1-\bar{\alpha}_t}} \hat{\boldsymbol{\epsilon}}_{t-1}\Big) + \sigma_t \boldsymbol{\epsilon},
\end{equation}
where $\sigma_t$ denotes the standard deviation of the stochastic term. In practice, both $\sigma_t = \sqrt{\beta_t}$ and $\sigma_t = \sqrt{\frac{1-\bar{\alpha}_{t-1}}{1-\bar{\alpha}_t}\beta_t}$ have been shown to yield comparable results. Following common practice, we adopt the simpler choice $\sigma_t = \sqrt{\beta_t}$.

During training, a time step $t$ is uniformly sampled from $\{1,\dots,T\}$, and random Gaussian noise 
$\boldsymbol{\epsilon}_{t-1}$ is added to the latent state $\mathbf{z}_{t-1}$ to produce 
the noisy latent $\mathbf{z}_t$, as defined in \eqref{eqn:zt}. The optimal model parameters $\boldsymbol{\theta}^{\ast}_\text{diff}$ of the denoising network $\kappa(\cdot)$ are obtained by training the network to predict the added noise, minimizing the estimation error:
\begin{equation} 
\boldsymbol{\theta}^{\ast}_\text{diff} =  \operatorname*{arg\,min}_{\boldsymbol{\theta}_\text{diff}} \vert \kappa_{\boldsymbol{\theta}_\text{diff}}(\mathbf{z}_t, \mathbf{z}_c, t)
 - \boldsymbol{\epsilon}_{t-1} \vert^2.
\end{equation}
In practice, the loss sums over mini-batches sampled uniformly across all time steps, and training proceeds until convergence.

During inference, a noise latent vector $\mathbf{z}_T$ is sampled from Gaussian noise and progressively denoised using the trained conditional diffusion model, guided by the latent representation of an observed TVIR and the current step $t$. After $T$ steps, the resulting denoised latent vector $\mathbf{z}_0$ is decoded by the decoder $\mathcal{D}(\cdot)$ \correct{of the autoencoder} to recover the amplitude and sine and cosine components of the phase, which are then combined to reconstruct the $20 \times 250$ complex-valued TVIR in the original signal space.

\section{Model Training}
\subsection{Pre-training on Simulated Dataset}
Data-driven ML models, such as LSTM autoencoders and diffusion models, typically require large amounts of high-quality training data. In underwater environments, however, collecting extensive experimental datasets with high signal-to-noise ratios~(SNR) is particularly challenging. To address this limitation, we first pre-train the LSTM autoencoder and diffusion model on a large corpus of simulated TVIRs generated under diverse environmental conditions. This pre-training stage enables the models to learn fundamental characteristics of underwater acoustic propagation, such as multipath structure, Doppler effects, and \new{amplitude-phase} relationships, before being fine-tuned on real-world measurements.

We generated one million simulated TVIRs by randomly sampling key environmental parameters according to the distributions listed below:
\begin{table}[ht]
\centering
\small
\renewcommand{\arraystretch}{1.2}
\begin{tabular}{|l|l|}
\hline
\textbf{Environmental Parameter} & \textbf{Distribution} \\
\hline
Surface sound speed & $\mathcal{U}(1500, 1550)$~m/s \\
\hline
Sound speed gradient & $\mathcal{N}(0,\,0.05)$~1/s \\
\hline
Water depth~($d$) & $\mathcal{U}(10, 100)$~m \\
\hline
Source depth & $\mathcal{U}(2.5, d-2.5)$~m \\
\hline
Receiver depth & $\mathcal{U}(2.5, d-2.5)$~m \\
\hline
\new{Source-receiver} range & $\mathcal{U}(1, 1000)$~m \\
\hline
Relative density& $\mathcal{U}(1.145, 2.500)$ \\
\hline
Relative sound speed & $\mathcal{U}(0.98, 2.50)$ \\
\hline
Absorption & $\mathcal{U}(0.0000, 0.0022)$ \\
\hline
Surface reflection coefficient & $\mathcal{U}(0.5, 1.0)$ \\
\hline
\end{tabular}
\caption{Ranges and distributions of environmental parameters used for simulated TVIR generation. $\mathcal{U}$ denotes a uniform distribution, and $\mathcal{N}$ denotes a normal distribution.}
\label{tab:env_params}
\end{table}

\begin{figure}[h]
	\centering
	\includegraphics[width=0.45\textwidth]{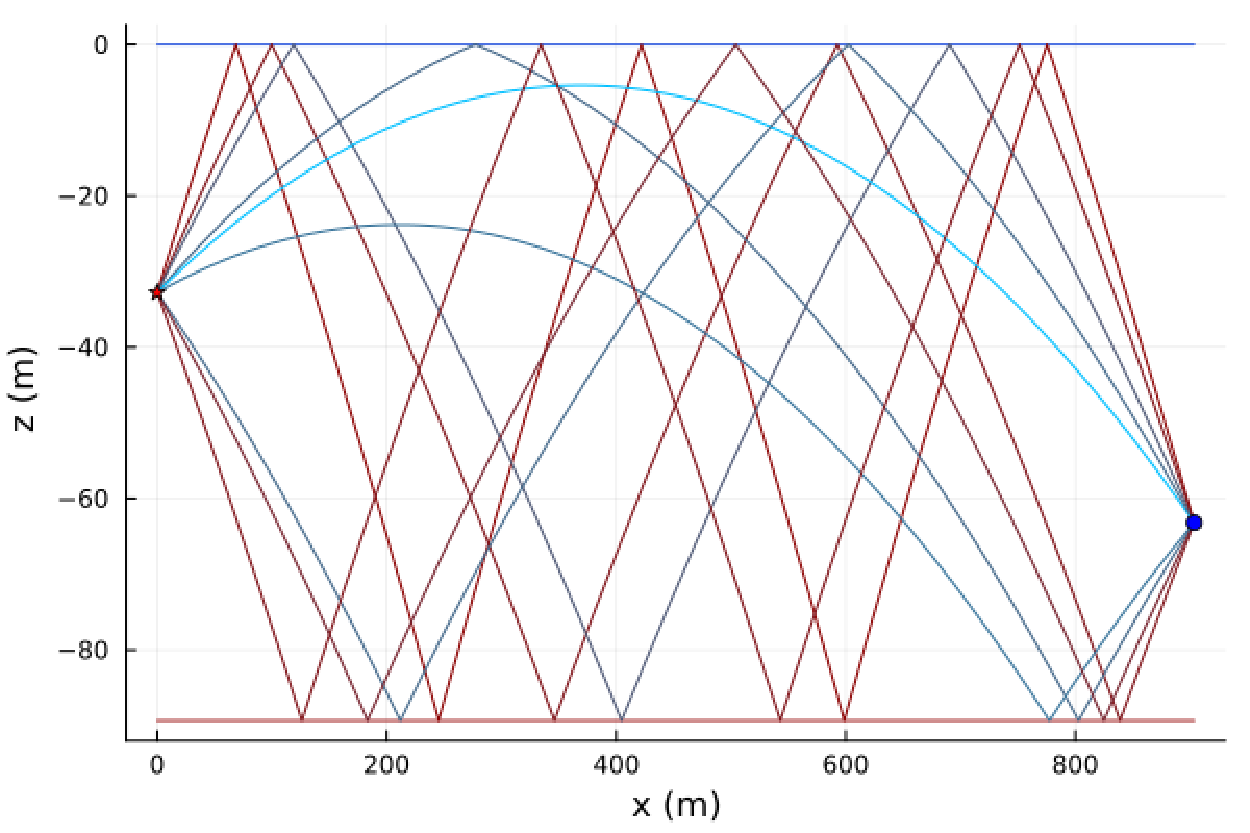}
	\caption{Ray trajectories computed by \texttt{RaySolver} for a representative simulated underwater environment. The red star indicates the source location, and the blue dot indicates the receiver location.}
	\label{fig:env}
\end{figure}

\begin{figure*}[h!]
	\centering
	\begin{subfigure}{0.45\textwidth}
\includegraphics[width=\textwidth]{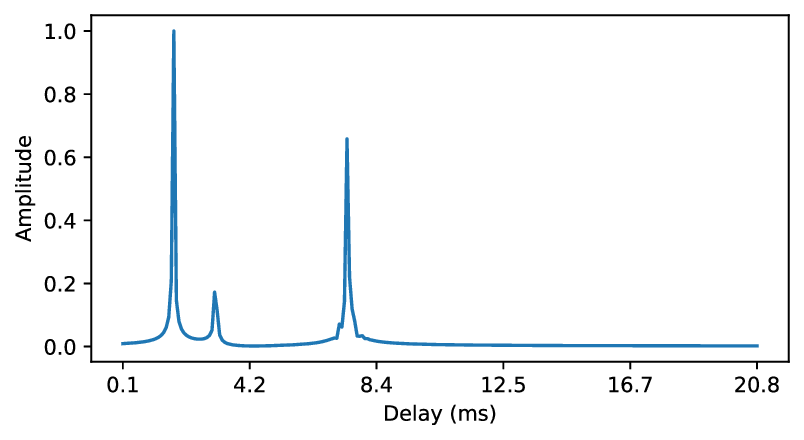}
 	\caption{Nominal CIR.}
 \end{subfigure}
  \begin{subfigure}{0.45\textwidth}
 \includegraphics[width=\textwidth]{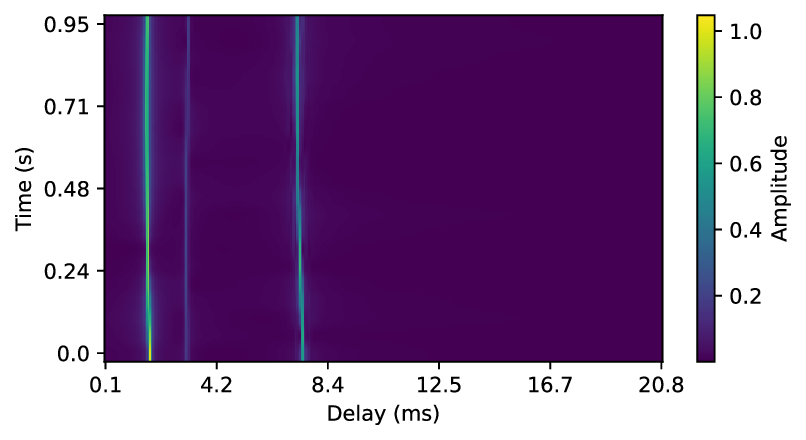}
 	\caption{Corresponding TVIR.}
 \end{subfigure}
\caption{Example of a simulated nominal CIR and the corresponding TVIR that generated by the \texttt{RaySolver} and the channel simulator.}
	\label{fig:example_tvir}
\end{figure*}

For each simulated environment defined by environmental parameters randomly sampled from the specified ranges, we used the \texttt{RaySolver} from the \texttt{UnderwaterAcoustics.jl} package~\cite{UnderwaterAcoustics,chitre2023ua} to generate a nominal CIR. Fig.~\ref{fig:env} illustrates the ray trajectories for a representative simulated underwater environment. To introduce temporal variations and obtain a full TVIR from the nominal CIR, we employed an acoustic channel simulator~\cite{qarabaqi2013acoustic}, which perturbs the nominal CIR using randomized, time-varying environmental parameters to emulate dynamic underwater conditions. \update{The resulting TVIR incorporates large-scale multipath structure together with small-scale time-frequency variations induced by Doppler effects, surface motion, and frequency-dependent absorption. The nominal CIRs are obtained from the \texttt{RaySolver}, while the acoustic channel simulator introduces stochastic perturbations to model small-scale fading and Doppler dynamics within the ranges specified in Table~\ref{tab:simulator_params}. This results in diverse TVIR realizations that capture variations in delay spread, scattering strength, and temporal dynamics.} An example nominal CIR generated by \texttt{RaySolver} and the corresponding TVIR produced by the simulator are shown in Fig.~\ref{fig:example_tvir}.

\begin{table}[ht]
\centering
\small
\renewcommand{\arraystretch}{1.2}
\begin{tabular}{|l|l|}
\hline
\textbf{Simulator Parameter} & \textbf{Value / Range} \\
\hline
Minimum frequency  & 18 kHz \\
\hline
Bandwidth & 12 kHz \\
\hline
Frequency resolution  & 24 Hz \\
\hline
Spreading factor ($k$) & 2 \\
\hline
Time resolution & 50 ms \\
\hline
Surface fluctuation strength  & 1.5 \\
\hline
Intra-path count  & 20 \\
\hline
Intra-path delay spread parameter  & $5 \times 10^{-4}$ \\
\hline
Doppler scaling factors  & $[1, 10)$ \\
\hline
Doppler fluctuation amplitudes & $a_3 \sim \mathcal{U}(0,2),\;$  \\
 & $a_4 \sim \mathcal{U}(0,1.5)$ \\
\hline
Wave amplitude & 0.05 \\
\hline
Wave frequency & 0.01 Hz \\
\hline
\end{tabular}
\caption{\update{Key configuration of the acoustic channel simulator for TVIR generation.}}
\label{tab:simulator_params}
\end{table}

\subsection{Autoencoder Performance}
\begin{figure}
\centering
\includegraphics[width=0.5\textwidth]{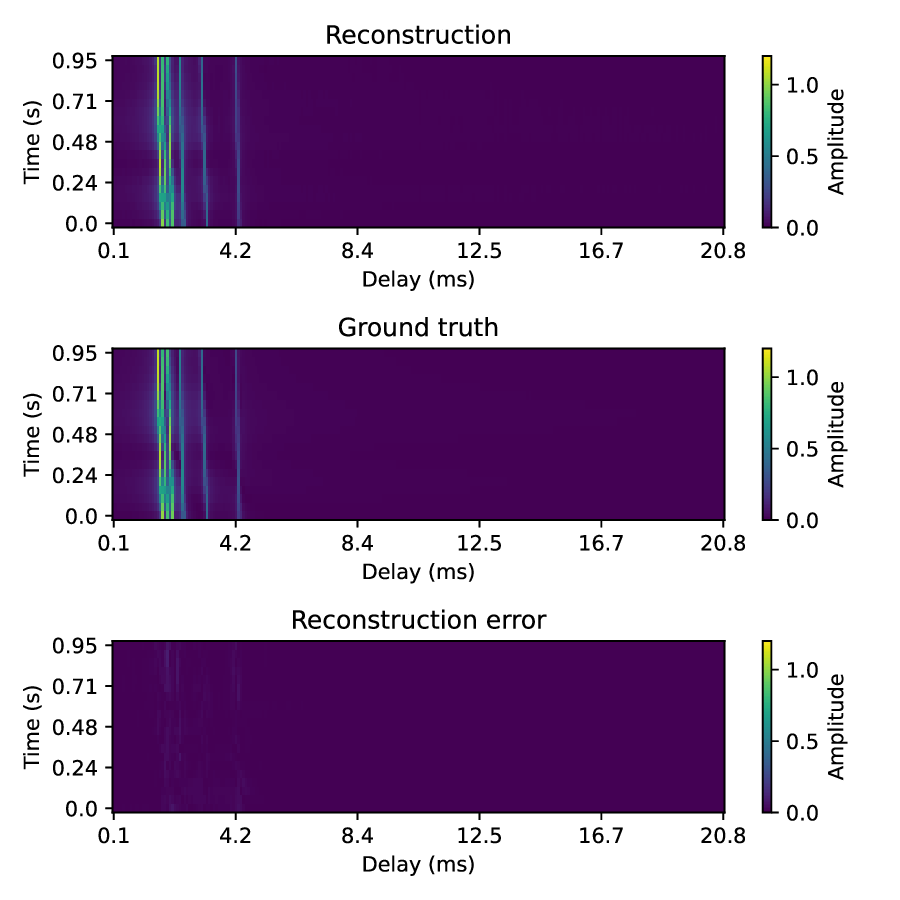}
\caption{A simulated test TVIR used to evaluate the reconstruction performance of the pre-trained LSTM autoencoder.}
\label{fig:AE_pretrain}
\end{figure}

We pre-trained the LSTM autoencoder on clean simulated TVIRs to capture essential channel characteristics. The training set comprised 900,000 simulated environments, with 99,500 reserved for validation and 500 for testing \update{with batch size of 256}. The initial learning rate was set to 0.001 and reduced by a factor of 10 if the validation error did not improve for three consecutive epochs. Early stopping was applied when the learning rate fell below $10^{-6}$. Training was performed using the ADAM optimizer~\cite{kingma2014adam} and typically completed within a few hours on an NVIDIA RTX A6000 GPU. Fig.~\ref{fig:AE_pretrain} presents the reconstruction results on the clean test set, demonstrating that the autoencoder effectively captures key amplitude variations and generalizes well to unseen environments under temporal variability.

\subsection{Diffusion Model Performance}
To pre-train the diffusion model, we first generated \correct{conditional} and target latent sequences using the pre-trained LSTM autoencoder. For each simulated TVIR, the channel simulator produced 40 consecutive CIRs over a 2~s duration from the nominal CIR, with the first 20 CIRs serving as the \correct{conditional} input and the remaining 20 forming the ground truth for the generated TVIR. To reduce computational cost, pre-training was formulated as a short-term prediction task rather than full generative modeling. We found that pre-training plays a more crucial role for the autoencoder than for the diffusion model, as the latter is trained on a much smaller latent space. While training the diffusion model to generate entirely new sequences could better capture long-term environmental variations, focusing on short-term evolution enabled efficient learning of key channel features and temporal correlations across diverse environments. With additional computational resources, extending pre-training to the full generative task could further improve model performance by capturing long-term dependencies and a broader range of channel variations.

\begin{figure}
\centering
\includegraphics[width=0.5\textwidth]{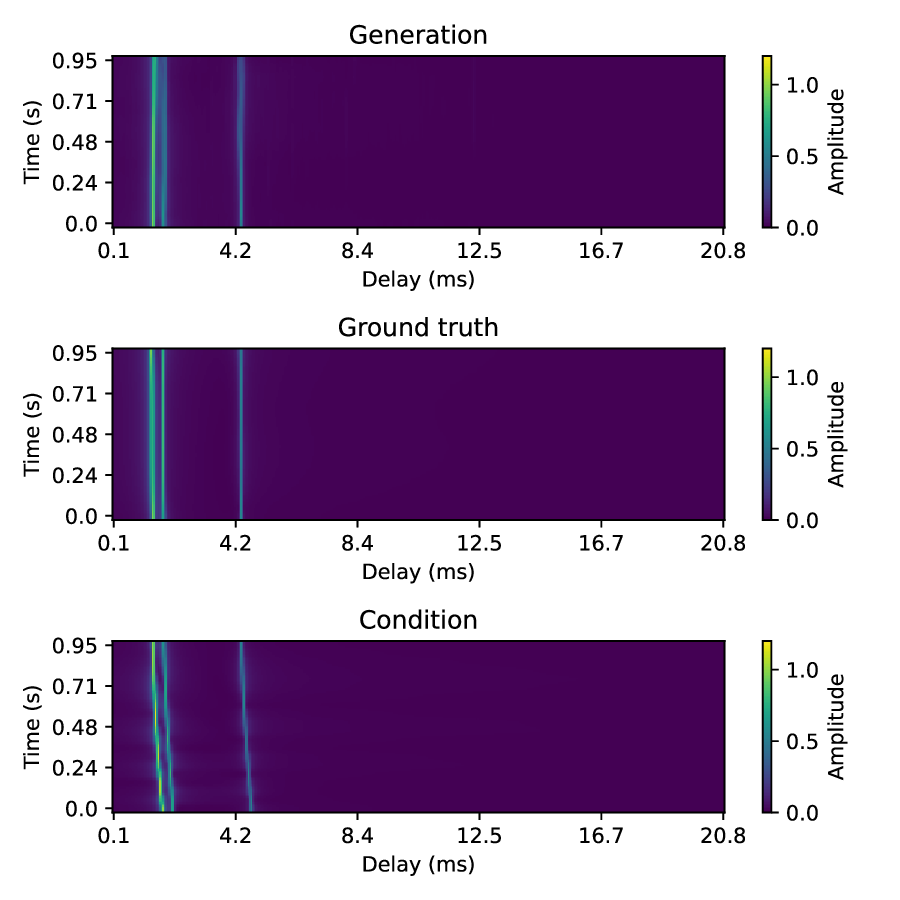}
\caption{A simulated test TVIR used to evaluate the performance of the pre-trained conditional latent diffusion model.}
\label{fig:diff_pretrain}
\end{figure}

The \correct{conditional} and target TVIRs were encoded into latent space using the pre-trained LSTM autoencoder. The diffusion model was then trained to predict latent sequences that, when decoded, reproduced TVIRs consistent with the target sequences. A total of 900,000 \new{\correct{conditional}-ground} truth latent pairs were used for training, with 99,500 reserved for validation and 500 for testing. \update{The batch size is set as 256}. Training began with a learning rate of 0.001, which was reduced by a factor of 10 if the validation loss did not improve for five consecutive epochs. Early stopping was applied once no further improvement was observed and the learning rate dropped below $10^{-6}$. The model employed 100 denoising steps with a linear noise schedule ranging from $10^{-4}$ to $10^{-2}$. Although alternative schedules have been explored in prior studies~\cite{chen2023importance,hang2024improved}, this simple configuration was used for pre-training, and the scheduler was treated as a tunable hyperparameter during fine-tuning, enabling task-specific optimization for improved denoising performance. Pre-training typically completed within one to two hours on an NVIDIA RTX A6000 GPU.

Fig.~\ref{fig:diff_pretrain} shows an example TVIR generated by the pre-trained diffusion model from simulated data. In most cases, the model produced realistic extensions of the conditional latent sequences. When decoded back to signal space, the generated TVIRs remained physically plausible, though small deviations from the ground truth were occasionally observed due to the stochastic nature of sampling and the model's approximation of the underlying statistical distribution.

\subsection{Experimental Fine-tuning}

In November 2024, we conducted a preliminary experiment to record channel soundings using mid-frequency (18--30~kHz band) modems in Singapore. The collected data provided a real-world basis for testing and fine-tuning our pre-trained models. During the experiment, the receiver modem was stationed at the jetty, while the transmitter modem was mounted on a boat anchored at various locations within the bay and allowed to drift while continuously transmitting. Channel soundings employed 8191-bit m-sequences, repeated five times, with a 24~kHz sampling rate and 24~kHz center frequency. Multiple consecutive transmissions were recorded at each location, resulting in a total of 28 recordings\footnote{\update{In this work, a ``recording" denotes a raw sea-trial measurement before preprocessing, whereas a ``TVIR" refers to the processed and segmented representation derived from each recording for model training and evaluation.}}, each approximately one minute long.

\begin{figure}[h!]
	\centering
\begin{subfigure}{0.45\textwidth}
 \includegraphics[width=\textwidth]{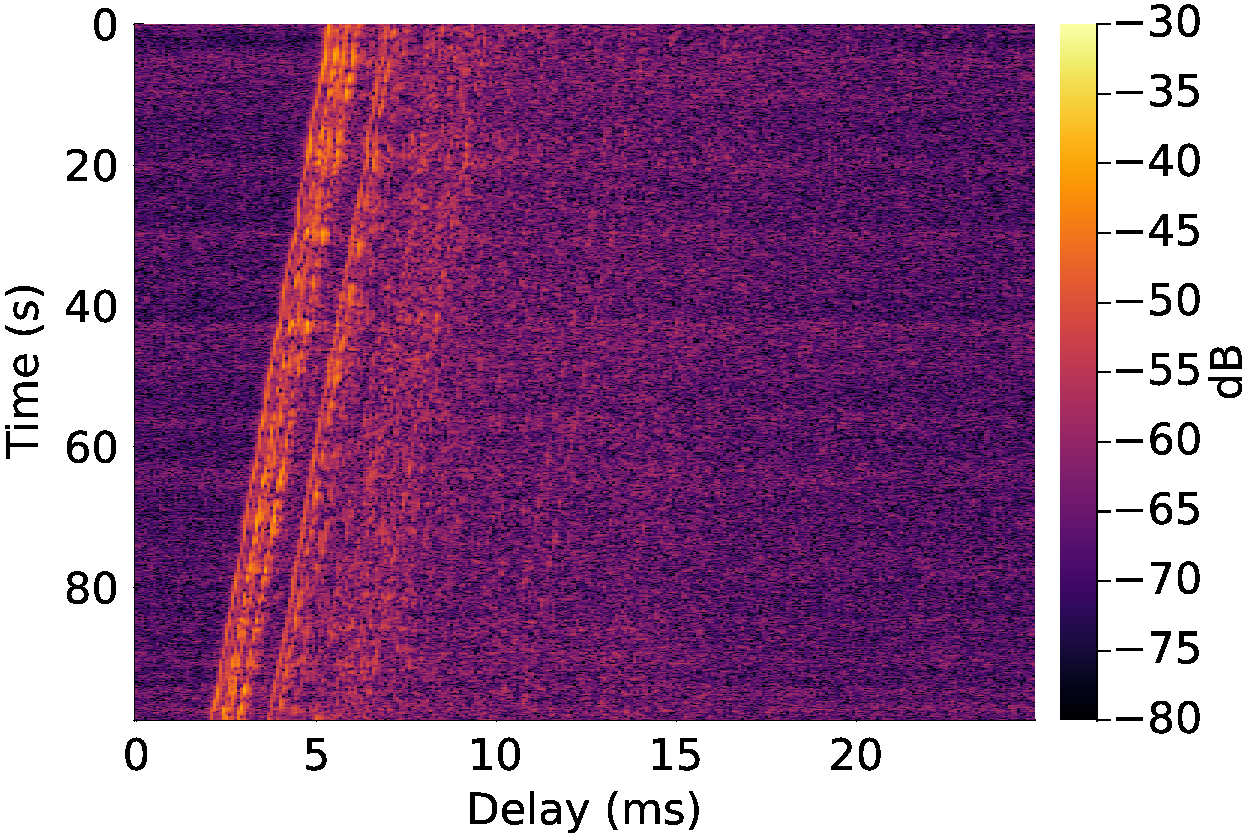}
 	\caption{Recording 1.}
 \end{subfigure}
 \begin{subfigure}{0.45\textwidth}
 \includegraphics[width=\textwidth]{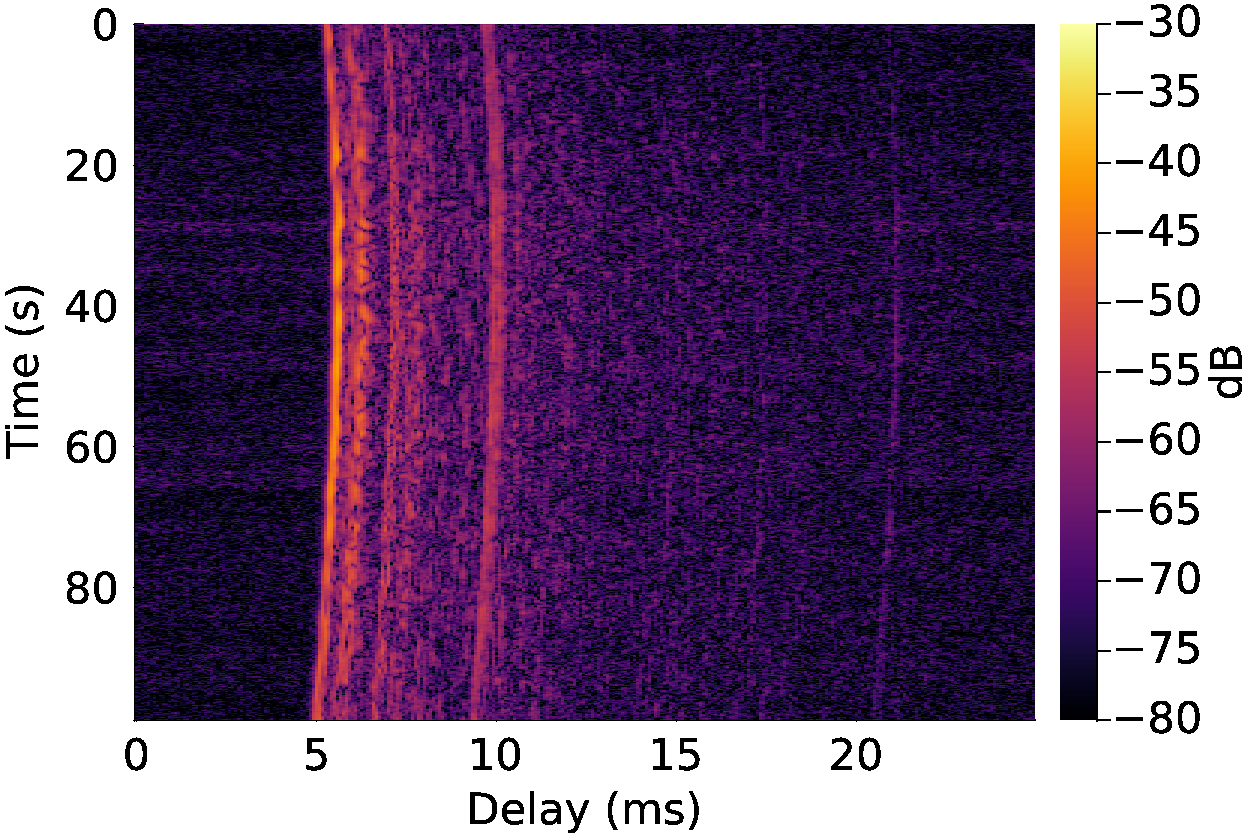}
 	\caption{Recording 2.}
 \end{subfigure}
 \caption{Example channel responses estimated from two recordings at Singapore.}
	\label{fig:first_trial_recording}
\end{figure}

\begin{figure}[h!]
	\centering
	\begin{subfigure}{0.45\textwidth}
 \includegraphics[width=\textwidth]{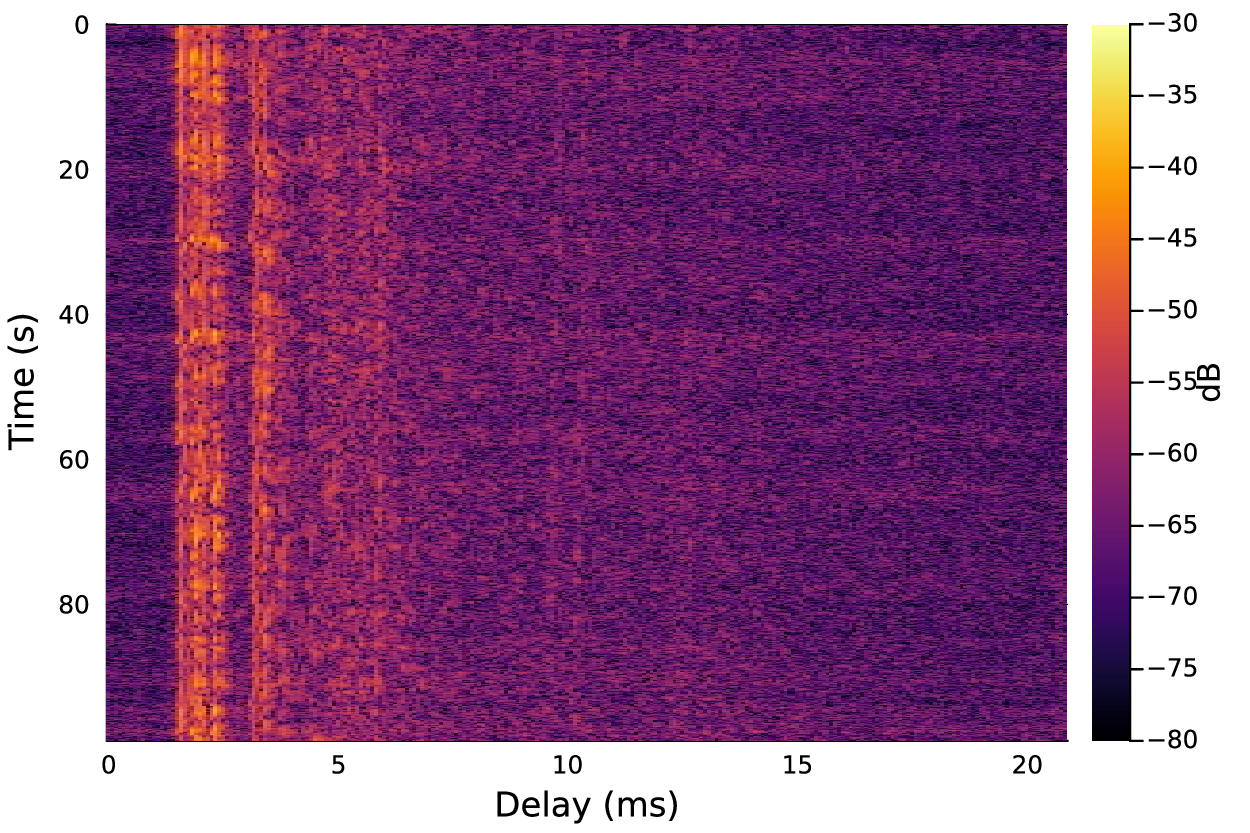}
 	\caption{Recording 1.}
 \end{subfigure}
  \begin{subfigure}{0.45\textwidth}
 \includegraphics[width=\textwidth]{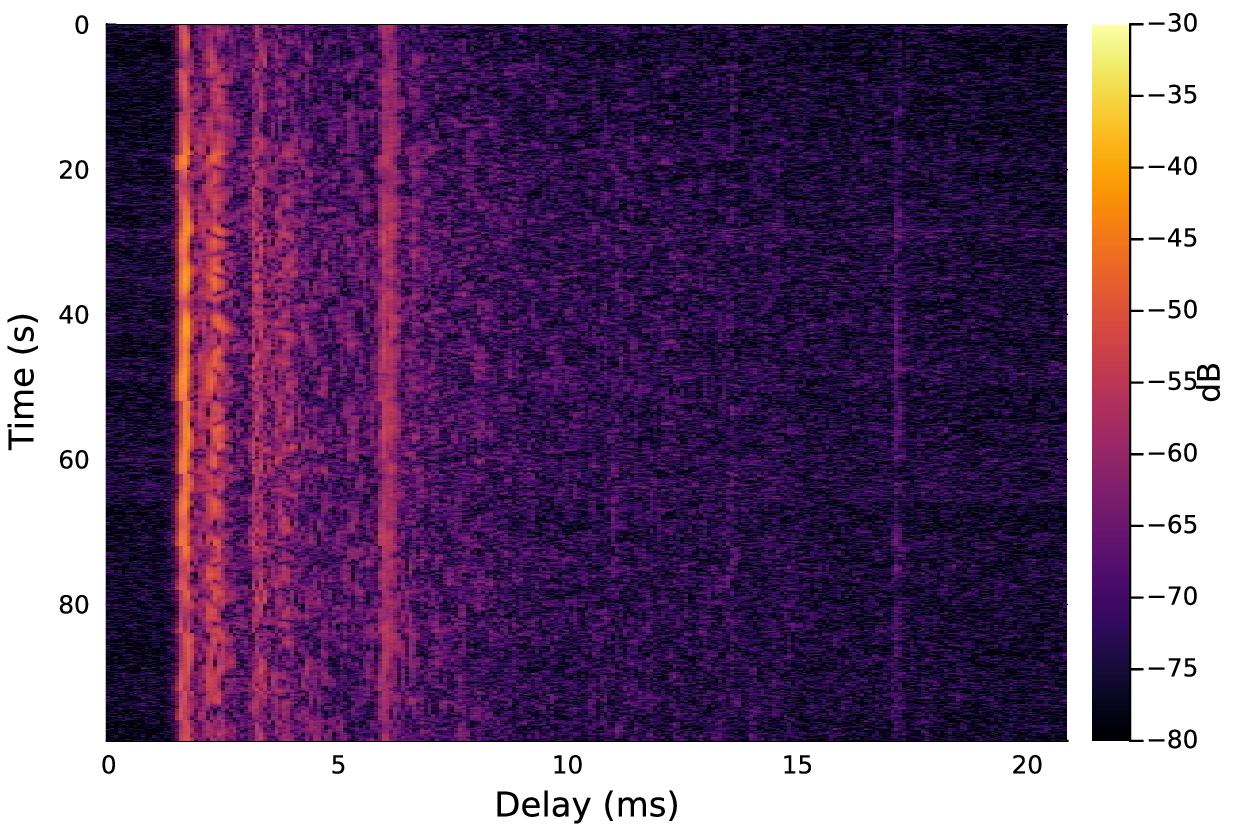}
 	\caption{Recording 2.}
 \end{subfigure}
\caption{\update{Illustration of the normalized recordings from Fig.~\ref{fig:first_trial_recording}. For visualization purposes, each TVIR is aligned based on the dominant arrival component of the corresponding CIR, providing a consistent view of the temporal structure across recordings.}}
	\label{fig:straightening_tvir}
\end{figure}

Measured TVIRs were extracted from the raw recordings using the normalized least-mean-squares~(NLMS) adaptive filtering algorithm~\cite{haykin2002adaptive} implemented in \texttt{AdaptiveEstimators.jl}~\cite{adaptiveestimation1}. Fig.~\ref{fig:first_trial_recording} shows two example channel responses from different recordings, illustrating delay variations caused by the drifting transmitter. To facilitate model training, amplitude and index normalization was applied to align the recordings. This straightening preserved the spatio-temporal evolution of the channel taps within the TVIR window of interest while providing the model with consistently aligned inputs that were easier to learn from. Fig.~\ref{fig:straightening_tvir} illustrates this procedure by stacking straightened TVIRs from consecutive recordings, demonstrating how the normalization removed delay drift while retaining the temporal dynamics of the channel.
 
The measured channels in Singapore differed from our clean simulated data primarily due to real-world background noise. To improve model robustness, we first fine-tuned the pre-trained models using augmented noisy simulated data, where the noise was modeled as Gaussian with a correlation structure estimated from the experimental measurements. This allowed the models to handle background noise effectively. \update{To augment the simulated data, we construct an empirical noise model from 28 experimental recordings. Each recording is segmented into TVIRs, and noise samples are aligned with respect to the direct arrival index. For each delay index, a 20-dimensional noise vector is collected across recordings to form an empirical distribution of noise statistics. This distribution is approximated using a multivariate Gaussian model to capture the second-order statistics of the noise at each delay. During augmentation, noise vectors are sampled from the corresponding distribution and added to the clean TVIR at each delay index. A different recording-based noise profile is randomly selected for each TVIR to ensure diverse realizations across the dataset. This procedure generates an augmented dataset that reflects the statistical characteristics observed in real-world measurements. This empirical augmentation strategy enriches data variability and improves model robustness by serving as an intermediate step to bridge the gap between clean simulated TVIRs and real-world experimental measurements.}  For fine-tuning on the Singapore experimental data, 25 recordings were randomly selected for training, with 3 reserved for testing. CIRs were sampled at 20~Hz to construct one-second TVIR blocks, yielding 1,440 training TVIRs and 148 test TVIRs for the autoencoder.

\begin{figure}[h!]
\centering
\includegraphics[width=0.5\textwidth]{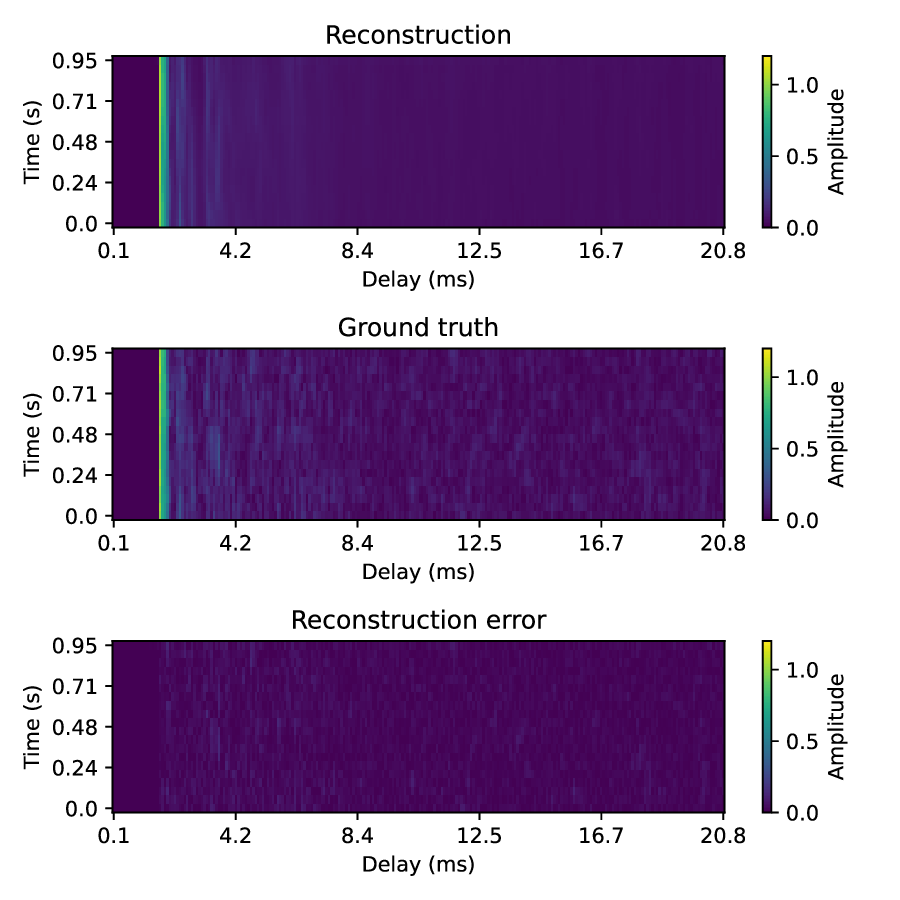}
\caption{An examples of TVIRs reconstructed by the experimentally fine-tuned autoencoder using sea-measured TVIRs collected from November 2024 experiment.}
	\label{fig:ae_exp}
\end{figure}

The autoencoder was fine-tuned with a small learning rate of 0.0001, reduced after five epochs without validation improvement. It was then further refined using the limited set of experimentally measured TVIRs to better adapt to real-world conditions. This experimental fine-tuning required roughly one minute on an NVIDIA RTX A6000 GPU, starting with a learning rate of 0.01 and a patience of 10 epochs. Training was terminated once the learning rate fell below $10^{-6}$, with ADAM used for optimization. Fig.~\ref{fig:ae_exp} shows the reconstruction performance of fine-tuned autoencoder on test dataset.

\begin{figure}
\centering
    \includegraphics[width=0.5\textwidth]{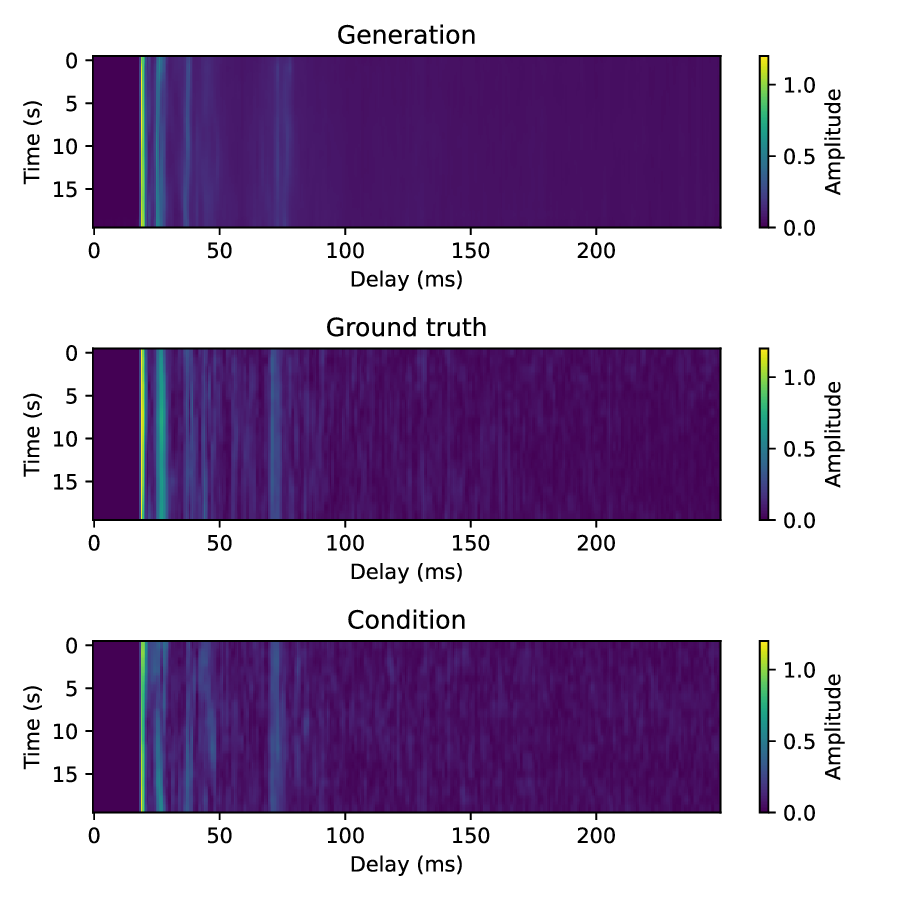}
	\caption{An examples of TVIRs generated by the experimentally fine-tuned diffusion model, conditioned on sea-measured TVIRs collected from November 2024 experiment.}
	\label{fig:diff_exp}
\end{figure}

The pre-trained diffusion model was refined using the latent representations of the experimentally collected data encoded by the experimentally fine-tuned encoder. Fig.~\ref{fig:diff_exp} illustrates the generation performance of the fine-tuned diffusion model on experimental test data from the November 2024 experiment. The generated latent vectors closely matched the \correct{conditional} inputs and, when decoded, preserved key arrivals in the original signal space, enabling reliable evaluation of candidate communication schemes using the generated channel responses.

\begin{figure*}
\centering
\includegraphics[width=\textwidth]{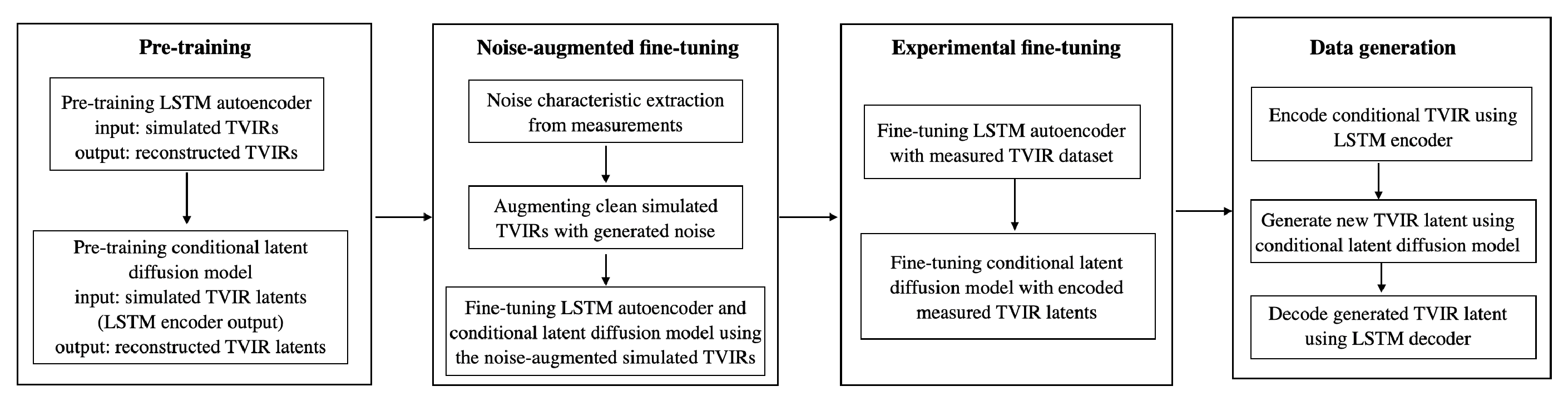}
\caption{\update{Overview of the full training and fine-tuning workflow for StableUASim.}}
	\label{fig:flow_summary}
\end{figure*}

The model, initially pre-trained on clean simulated TVIRs and subsequently fine-tuned on a limited set of experimental data, serves as our baseline before further adaptation to the environment of interest. Hereafter, we refer to it as the \emph{pre-trained baseline} model. Despite the small amount of experimental data used for pre-training, it effectively captures key arrivals and characteristic channel features of shallow Singapore waters, generating realistic responses for unseen TVIRs without additional adaptation. As a preliminary study, the current baseline demonstrates promising robustness and reliability, and its generalization performance is expected to improve further with a larger experimental dataset for pre-training. \update{Fig.~\ref{fig:flow_summary} summarizes the overall workflow of our approach, including pre-training, noise-augmented fine-tuning, experimental fine-tuning for rapid adaptation to the target environmental characteristics, and final TVIR generation using the locally fine-tuned models.}

\section{Experimental Validation}

To evaluate the proposed channel surrogate model, we consider two real-world underwater acoustic datasets and compare the model's performance against state-of-the-art methods. The first dataset, \textit{NOF1}, consists of shallow-water measurements from the publicly available \textit{Watermark}~\cite{7932436} dataset, collected in Norway Oslofjord. The second dataset comprises TVIRs recorded in a harbour near Keppel Marina, Singapore. \update{For both datasets, we adopt a conditional channel generation setting, where condition-ground truth pairs are dynamically resampled at each training epoch from the full measurement set. This enables the model to learn the underlying channel distribution rather than fixed one-to-one mappings. A separate held-out set of condition-ground truth pairs is used for validation and early stopping.} Our evaluation focuses on key communication metrics, particularly the bit error rate~(BER) as a function of SNR, complemented by statistical and structural analyses of the channels.

\subsection{Watermark Norway Oslofjord~(NOF1) Channel Dataset}
\label{sec:NOF}

\begin{figure}[h!]
\centering
    \includegraphics[width=0.5\textwidth]{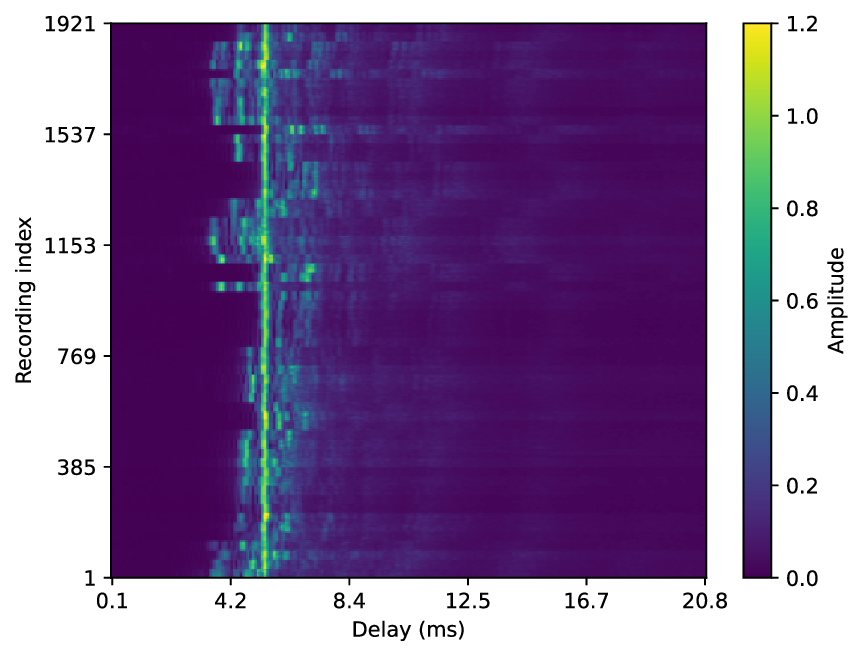}
    \caption{Amplitude profile of measured TVIRs in the complete NOF1 dataset.}
    \label{fig:NOF_amp}
\end{figure}

The NOF1 dataset represents a shallow-water acoustic communication scenario in Norway Oslofjord and is part of the Watermark measurement suite. Measurements were conducted in June in a fjord environment, with a transmission range of approximately 750~m and a water depth of about 10~m. The setup employed a stationary, bottom-mounted transmitter and receiver in a single-input single-output configuration. A linear frequency-modulated pulse train spanning 10--18~kHz was used as the probe signal. Each sounding lasted 32.9~s, with a Doppler spread of up to 7.8~Hz. A total of 60 independent TVIRs were recorded at 400~s intervals over roughly 33 minutes. Fig.~\ref{fig:NOF_amp} shows amplitude variation across the entire NOF1 dataset. Overall, the channels in NOF1 are relatively mild and stable~\cite{7932436}.

Our target application involves transmitted signals of approximately one-second duration, and StableUASim is designed accordingly. The original NOF1 dataset contains 60 TVIR recordings, each 32.9~s long with 256 delay taps. To match our model's input size, each recording was divided into 32 non-overlapping one-second segments sampled at 8~Hz, and each segment was then upsampled to 20~Hz for StableUASim and truncated to 250 taps, yielding a total of 1920 TVIR samples for fine-tuning and conditioning. Similarly, UACC-GAN-generated channel realizations were segmented into one-second blocks to maintain a consistent evaluation setup. In contrast, one-second segments are too short for stochastic replay to fully capture both slow-fading and fast-fading dynamics; therefore, stochastic replay was applied to the full 32.9~s recordings first and then segmented into one-second blocks. To ensure consistent temporal resolution across methods, outputs from StableUASim were resampled from the native 20~Hz to the original NOF1 rate of 8~Hz. 

To adapt StableUASim to the target environment, we sequentially fine-tuned the pre-trained autoencoder and the conditional latent diffusion model using all available recordings, starting from the pre-trained baseline. The initial learning rate for both models was set to 0.005 and reduced by a factor of 10 whenever the validation loss plateaued. Early stopping was applied once the learning rate fell below $10^{-6}$, with the ADAM optimizer used for parameter updates. The patience parameter was set to 15 for the autoencoder and 50 for the diffusion model. \update{To fine-tune autoencoder, we use 1728 TVIRs are training dataset, 178 as validation dataset and the rest 20 TVIR as test dataset. The batch size is set as 128 for both autoencoder and conditional latent diffusion model. }

Diffusion model fine-tuning employed full generative training to improve accuracy. During this process, TVIRs in the training dataset were randomly paired, with one serving as the conditional input and the other as the target output. These pairs were reshuffled at each epoch to generate new \new{condition-ground truth} combinations, allowing the model to learn the underlying distribution. A fixed set of condition-generation pairs \update{consisting of 1920 TVIR pairs, matching the size of the training dataset} was maintained to monitor training error and guide early stopping.

This strategy ensures that the diffusion model learns to generate realistic TVIRs tailored to the target environment. We employed a sigmoid-shaped noise scheduler, which allows the model to first focus on fine-grained latent features during the early diffusion steps and progressively handle increasingly randomized latent states in later steps. The sigmoid-shaped noise schedule~\cite{guo2025comprehensive} is defined as:

\begin{equation}
\begin{aligned}
\beta_t &= 10^{-4} + \frac{10^{-2}-10^{-4}}{1 + \exp\Big(-10 \big(\frac{t}{T} - 0.5\big)\Big)}, \\
& \quad t=1,\dots,T.
\end{aligned}
\end{equation}
Although alternative noise schedules, numbers of diffusion steps, and network architectures could be used, these parameters were selected empirically to ensure stable training and high-quality generation for our specific dataset and application.

We benchmark the proposed StableUASim model against both state-of-the-art ML approaches and conventional stochastic replay methods. As a representative deep generative baseline, we include \emph{UACC-GAN}~\cite{10608448}, an underwater acoustic channel simulator that learns the statistical distribution of measured TVIRs and generates realistic channel realizations using a GAN. For the replay baseline, we adopt the widely used \emph{stochastic replay}~\cite{5898427} (referred to as trend replay in the UACC-GAN paper), which stochastically replays the fast-varying component while preserving the temporal variations observed in the measurements. These baselines are chosen because UACC-GAN represents the recent state-of-the-art, and its authors reported these methods as top-performing benchmarks. For fair comparison, we use the channel realizations released by the original UACC-GAN authors, ensuring faithful reproduction of their results.

\begin{figure}
    \centering
    \begin{subfigure}{0.24\textwidth}
        \includegraphics[width=\textwidth]{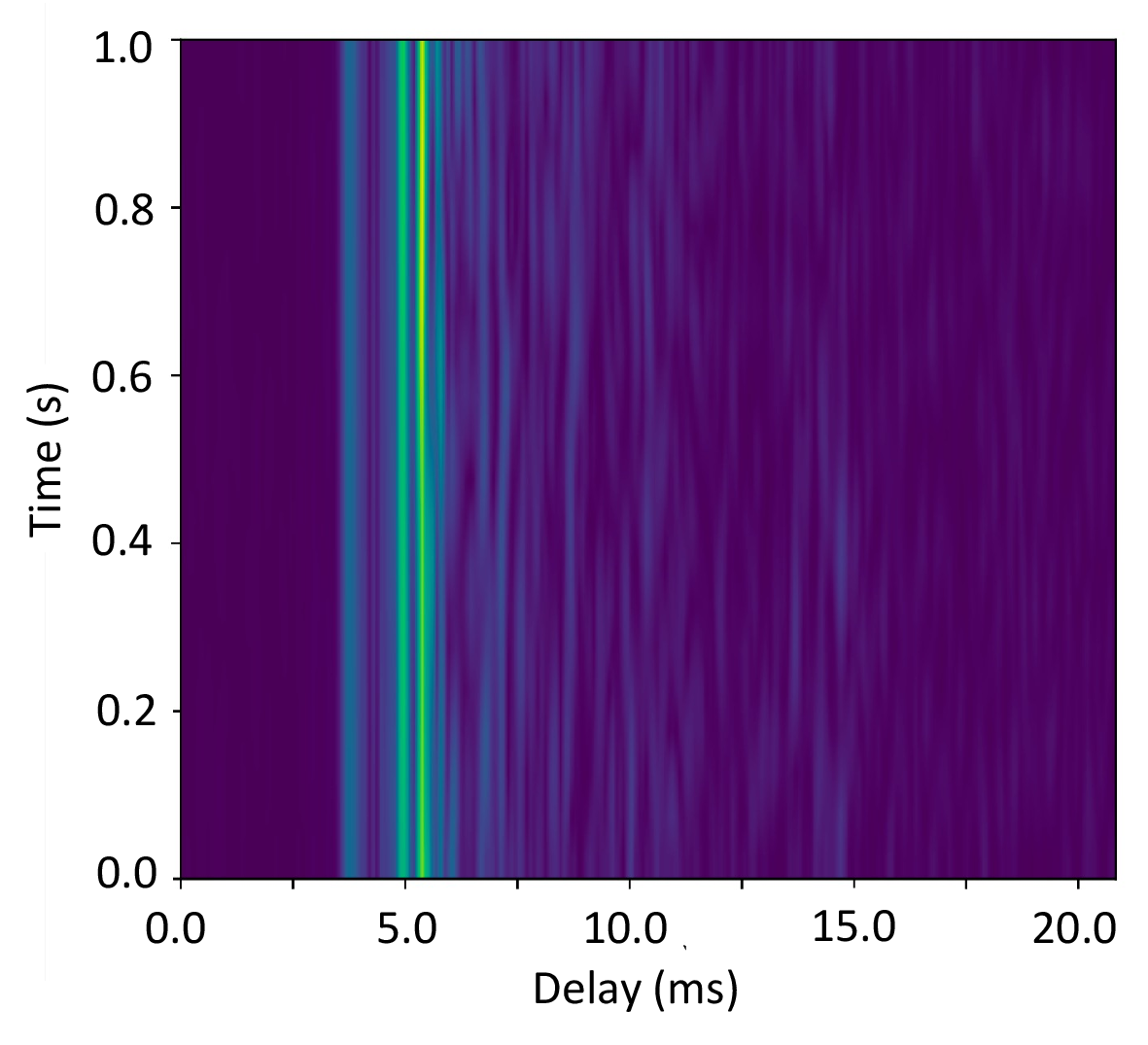}
        \caption{NOF1 measurement.}
    \end{subfigure}
    \begin{subfigure}{0.24\textwidth}
        \includegraphics[width=\textwidth]{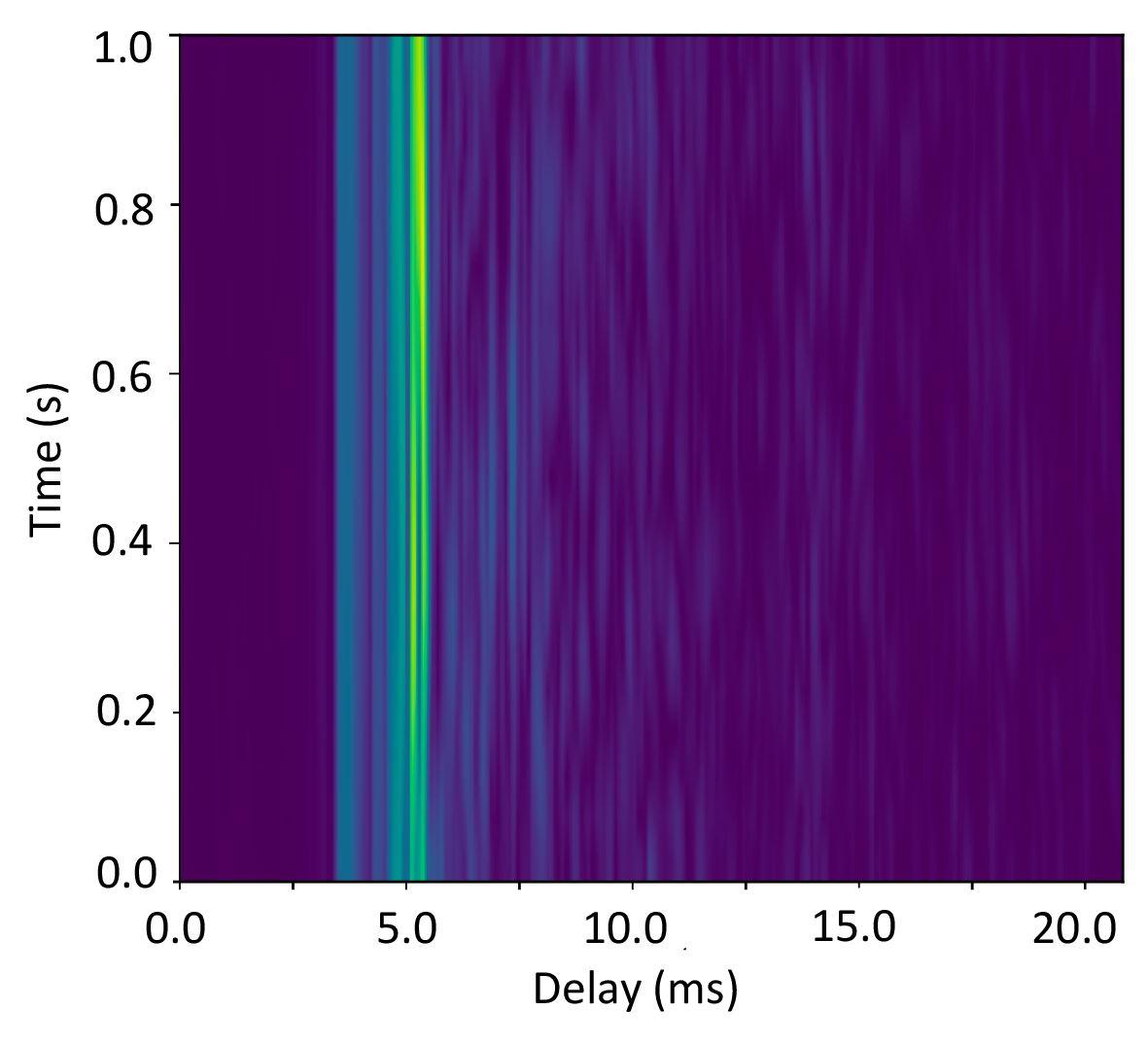}
        \caption{Stochastic replay.}
    \end{subfigure}
    \begin{subfigure}{0.24\textwidth}
        \includegraphics[width=\textwidth]{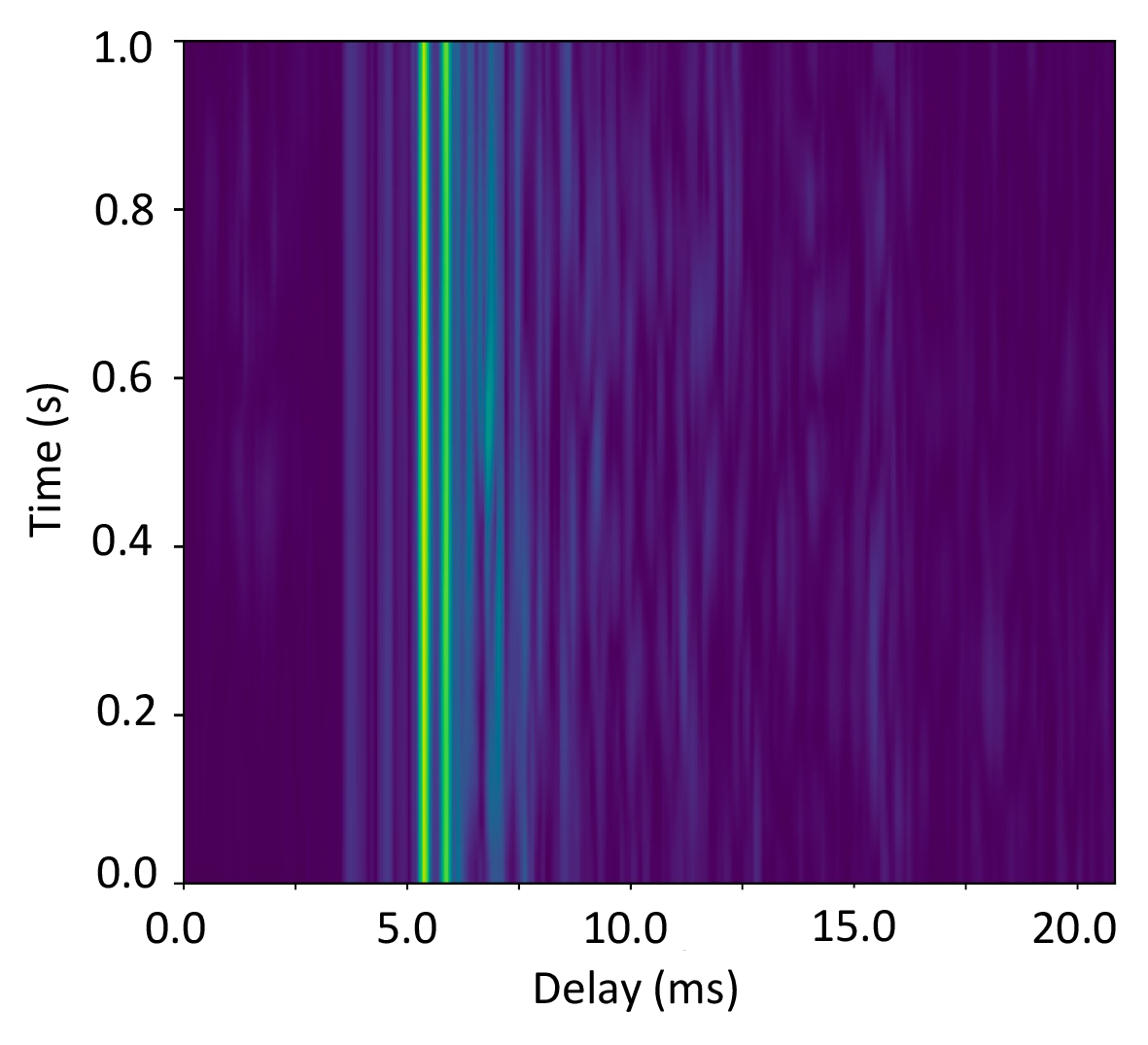}
        \caption{UACC-GAN.}
    \end{subfigure}
    \begin{subfigure}{0.24\textwidth}
        \includegraphics[width=\textwidth]{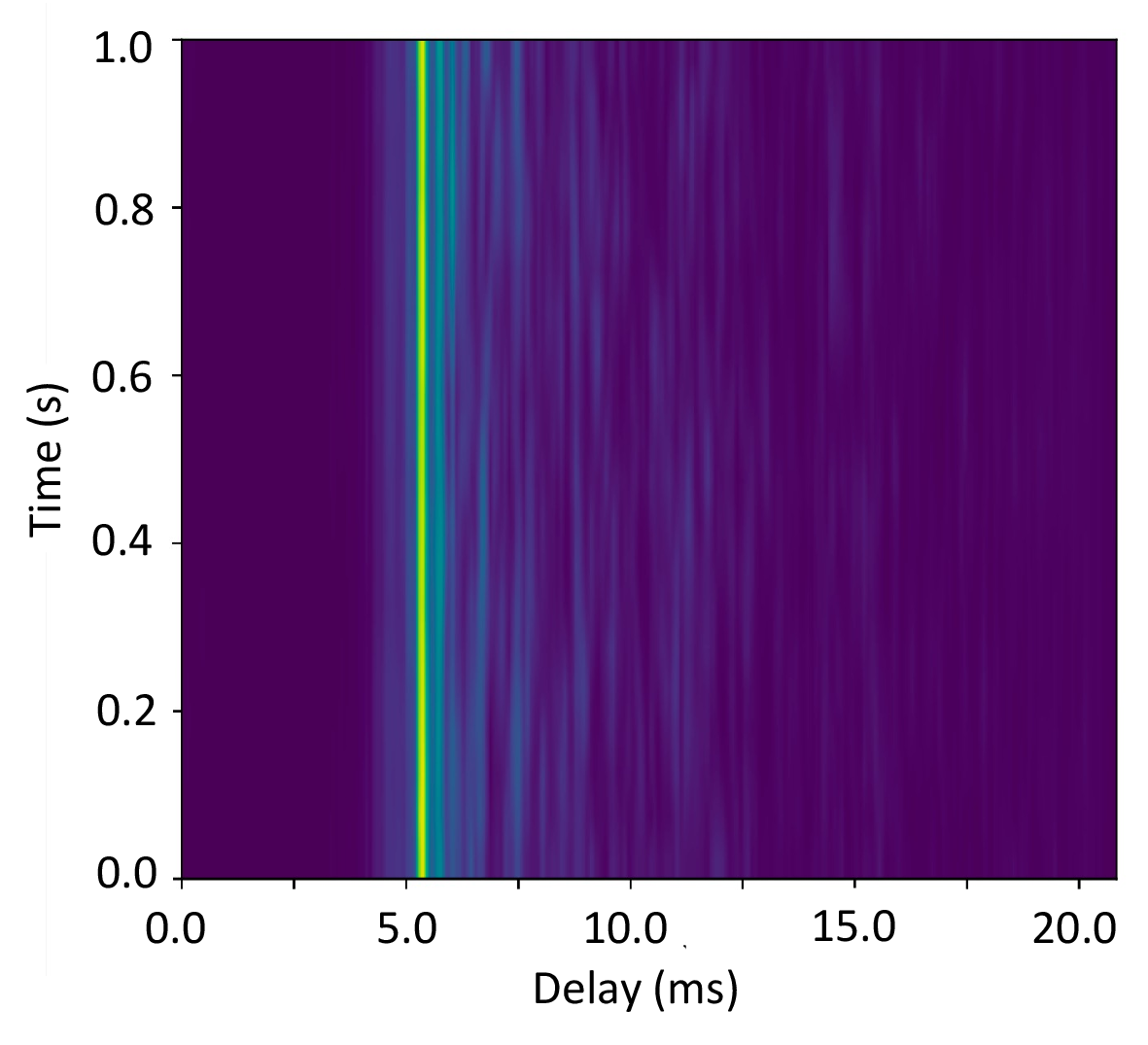}
        \caption{StableUASim.}
    \end{subfigure}
    \caption{Example TVIRs from the NOF1 measurement dataset and generated datasets for the three methods.}
    \label{fig:NOF_samples}
\end{figure}

\begin{figure*}[ht]
\centering
    \includegraphics[width=\textwidth]{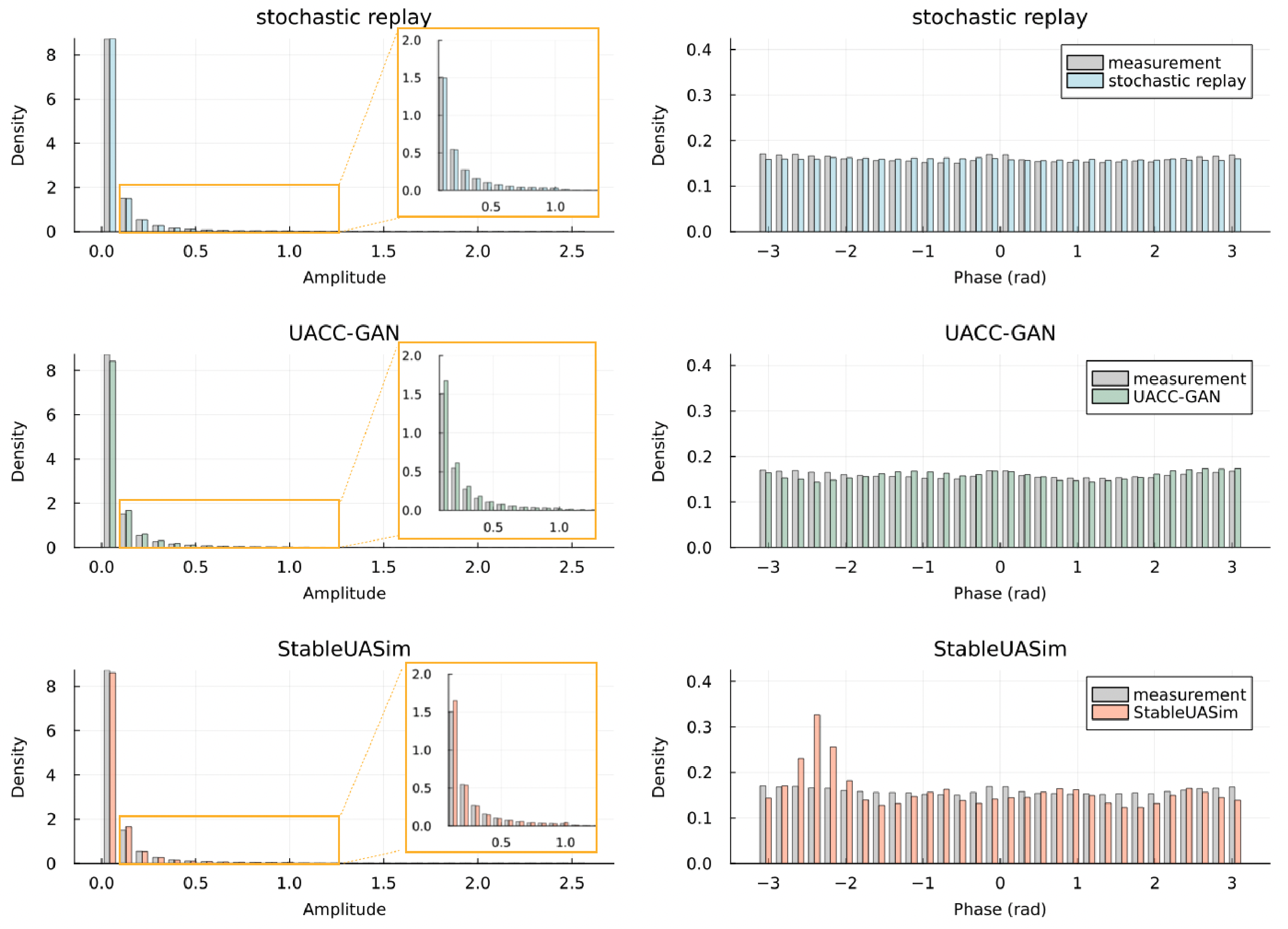}
    \caption{\update{Histograms of amplitudes and phases for all channel taps for NOF1 dataset, comparing generated channels with the measured dataset. The density is computed using normalized histograms, where counts are normalized by the total number of samples and bin width.}}
    \label{fig:NOF_histogram}
\end{figure*}

All channel datasets are normalized so that the maximum amplitude of the first CIR in each TVIR is unity, ensuring consistent comparisons across methods. Fig.~\ref{fig:NOF_samples} presents one representative measured TVIR alongside samples generated by each benchmarking model. The generated TVIRs are not in one-to-one correspondence with the measured instance as both UACC-GAN and our StableUASim model generate realistic samples drawn from the learned distribution rather than replicating specific measured instances. Visually, all three approaches produce responses qualitatively similar to the measured data. To examine the underlying structure, \update{we analyze the amplitude and phase distributions across all channel taps.} As shown in Fig.~\ref{fig:NOF_histogram}, all methods accurately capture the key distributions of both amplitude and phase. \update{We observe that the phase mismatch of StableUASim is primarily concentrated on very low-amplitude taps (below $-$26 dB), where the phase becomes less stable and exhibits noise-like behavior. This regime corresponds to weak multipath components that are dominated by random fluctuations, and thus contributes only marginally to the overall channel characteristics.}

\FloatBarrier
\begin{figure*}
    \includegraphics[width=\textwidth]{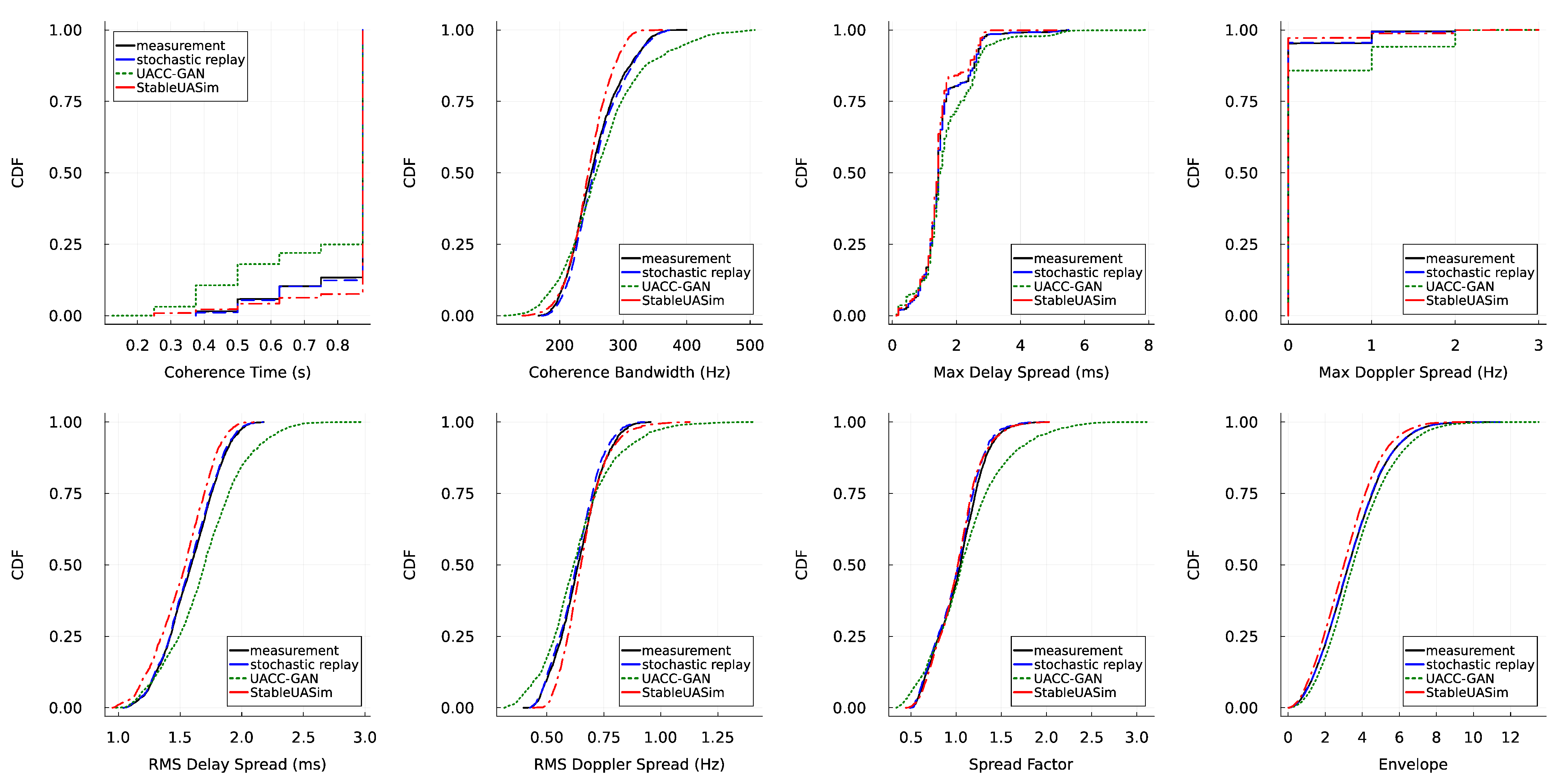}
    \caption{CDFs of eight key channel characteristics for the Watermark NOF1 dataset.}
    \label{fig:NOF_cdf}
\end{figure*}

\begin{figure}
\centering
    \includegraphics[width=0.5\textwidth]{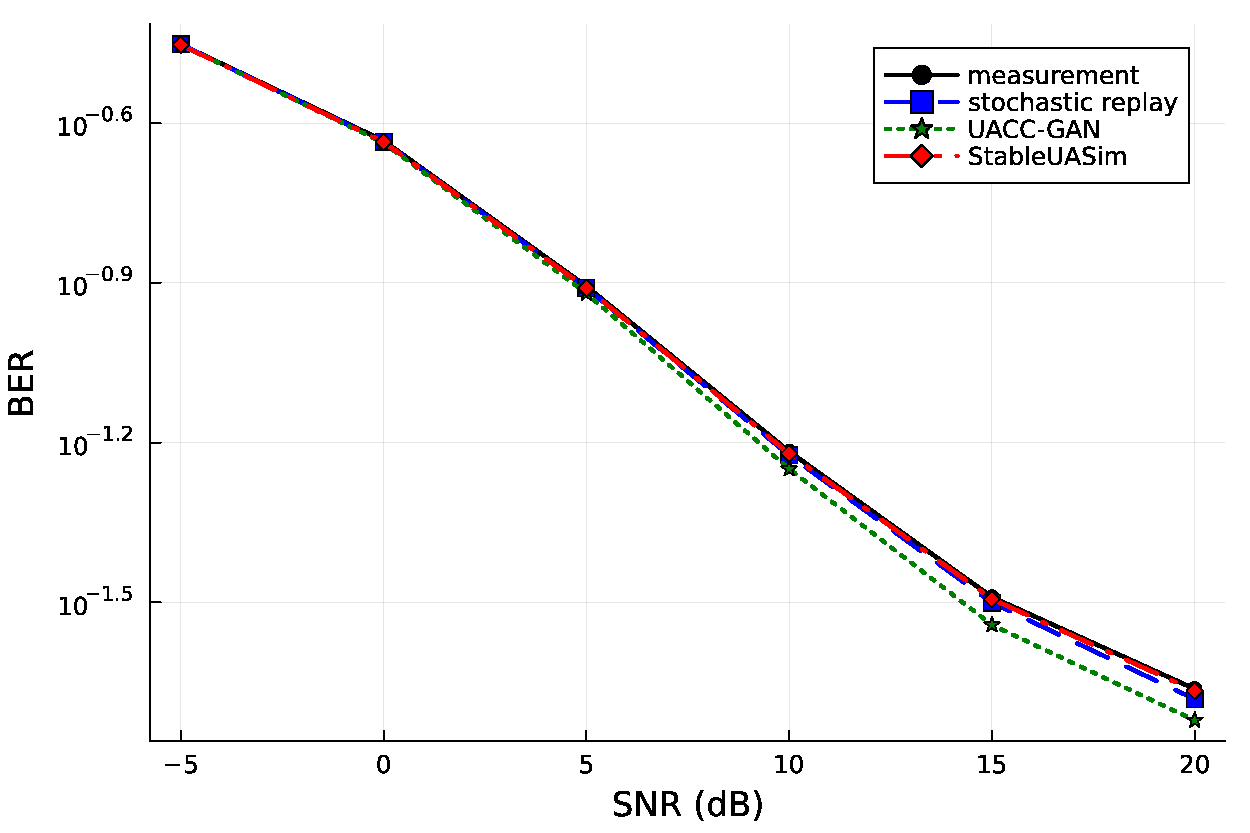}
    \caption{BER versus SNR for the Watermark NOF1 dataset.}
    \label{fig:NOF_ber}
\end{figure}

\begin{figure}
\centering
\begin{subfigure}{0.5\textwidth}
    \includegraphics[width=\textwidth]{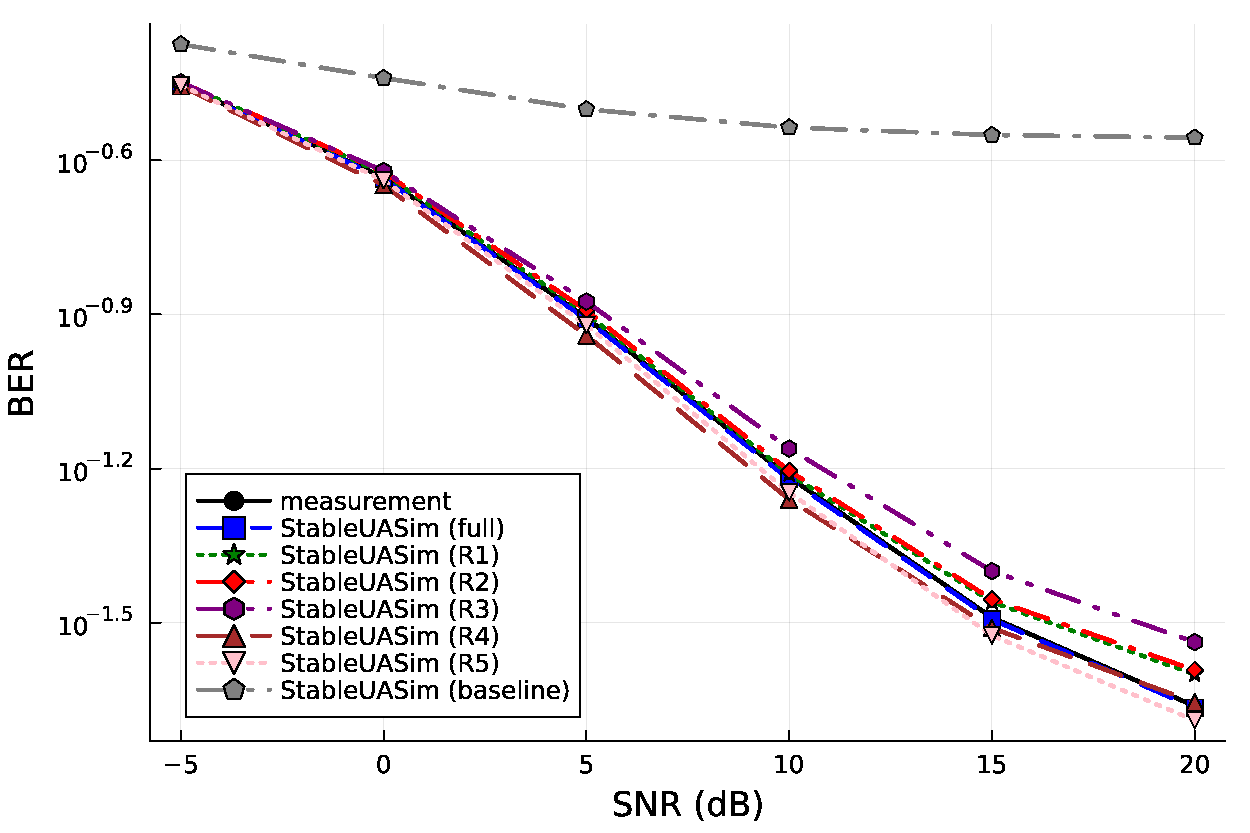}
    \caption{\update{BER curves across the full SNR range.}}
    \end{subfigure}
    \begin{subfigure}{0.49\textwidth}
    \includegraphics[width=\textwidth]{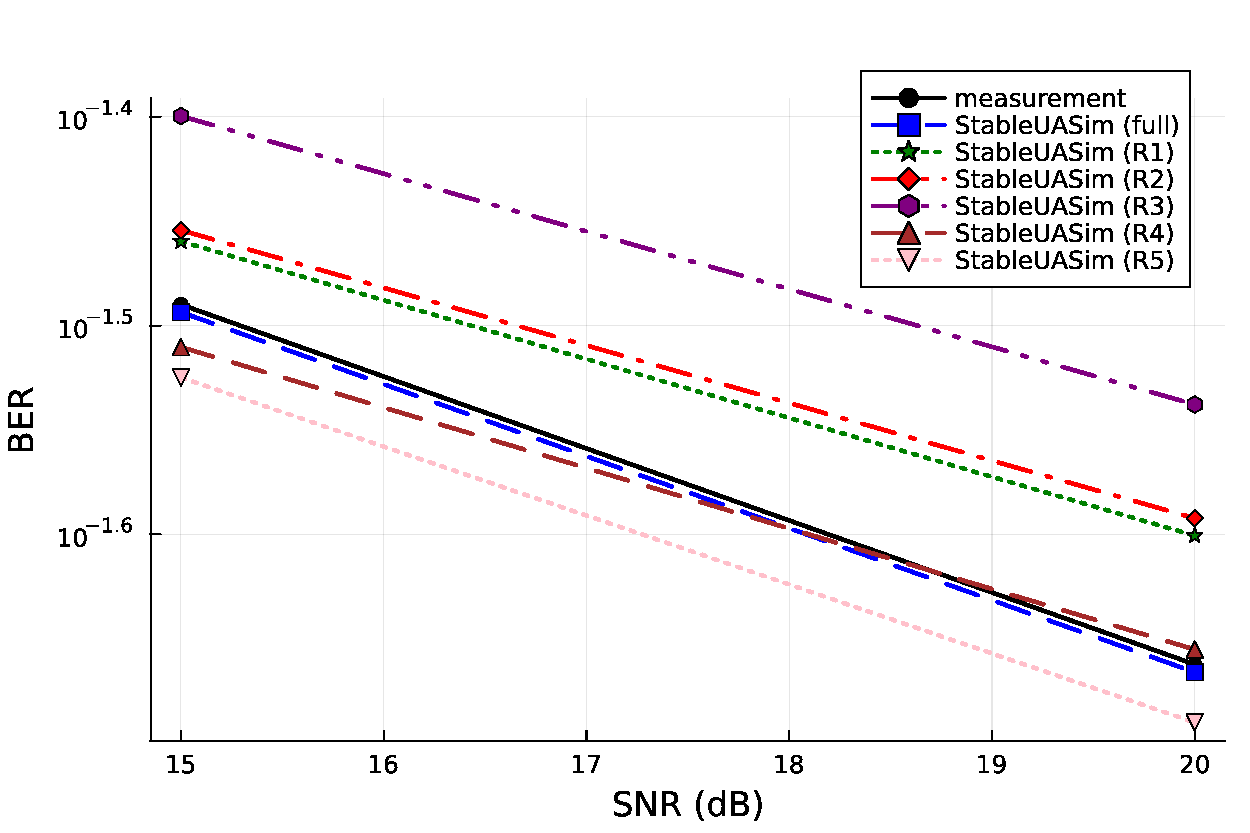}
    \caption{\update{Zoomed-in view of the high-SNR region, the baseline curve is omitted due to scale difference.}}
    \end{subfigure}
    \caption{\update{BER versus SNR for StableUASim fine-tuned on a single NOF1 recording across five runs~(R1-R5), where one recording is randomly selected in each run. The StableUASim model without any NOF1 fine-tuning is included as a baseline for comparison.}}
    \label{fig:NOF_ber_1rec}
\end{figure}

We evaluate communication performance using the same transmission scheme as in the UACC-GAN study. The signal spans 10--18~kHz and is transmitted via an OFDM system with 1024 subcarriers, 256 pilot symbols, a 0.04~s cyclic prefix, and QPSK modulation. No error-correction coding is applied to isolate the direct impact of the channel on communication performance without interference from coding schemes. The average BER is computed over all 1920 one-second TVIR samples for each SNR using an OFDM communication simulator. As shown in Fig.~\ref{fig:NOF_ber}, the BER curves from StableUASim closely match those from the measured channels, demonstrating accurate reproduction of real channel communication performance. We further compare the cumulative distribution functions~(CDFs) of eight key channel characteristics\footnote{The channel characteristic calculations follow the formulas provided in the UACC-GAN appendix~\cite{10608448}, ensuring consistency with the results reported in that work. The thresholds are set as follows: Doppler spread at -10~dB, delay spread at -10~dB, and coherence time at $\frac{\sqrt{2}}{2}.$} across the three modeling methods, using the measured dataset as reference. As illustrated in Fig.~\ref{fig:NOF_cdf}, our StableUASim model closely reproduces the statistical properties of the measured channels. \update{In particular, the close agreement in the CDF of channel characteristics and BER curves indicates that the dominant channel behavior is well captured by the proposed model, even though minor discrepancies may appear in very low-amplitude components, which are typically close to the measurement noise floor.}

The UACC-GAN approach, as reported by its authors, failed to converge when trained with fewer than 25 NOF1 recordings. In contrast, our method employs a pre-training strategy to capture general acoustic channel characteristics, achieving significantly higher data efficiency and enabling rapid adaptation to unseen environments. To illustrate this, we fine-tuned the pre-trained StableUASim model using a single randomly selected NOF1 recording at a time, repeating the process five times with different recordings. Fig.~\ref{fig:NOF_ber_1rec} shows the resulting BER curves of the model fine-tuned in each run, illustrating the model's robustness to the choice of fine-tuning data. \update{A variant of StableUASim without NOF1 fine-tuning is included as a reference baseline to highlight the effectiveness of domain adaptation.} Even when fine-tuned on just one recording out of the total 60 used by UACC-GAN, StableUASim quickly adapts and achieves communication performance close to that of models trained on much larger datasets, where UACC-GAN fails to function.

\subsection{Keppel Marina Dataset}

To further validate the proposed StableUASim model, we conducted measurements in Singapore waters at Keppel Marina. The objective was to evaluate the ability of both the stochastic replay approach and StableUASim to predict underwater communication performance in a controlled, realistic, and slowly varying environment. We do not include UACC-GAN in this comparison, as its architecture is specifically tailored to the Watermark dataset channels. Adapting UACC-GAN to our shorter TVIR sequences would require structural modifications that could compromise its optimization and yield an unfair comparison. Consequently, we focus exclusively on benchmarking StableUASim against the replay methods using this experimental dataset.
\begin{figure}
	\centering
	\includegraphics[width=\linewidth]{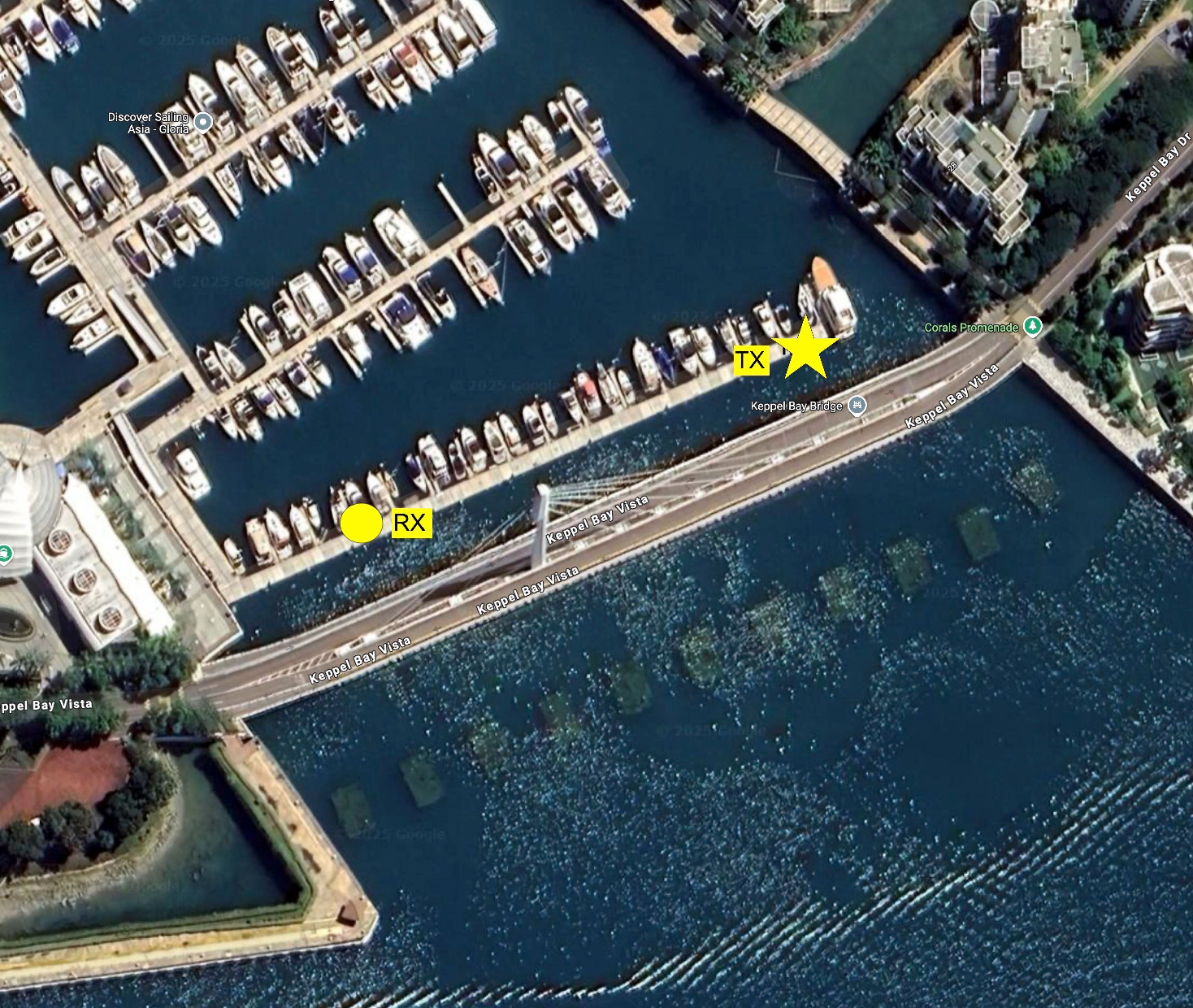}
	\caption{Illustration of the experimental setup at Keppel Marina. The transmitter~(TX) is surface-deployed, and the receiver array~(RX) is bottom-mounted, with hydrophones spaced within 30~cm of each other.}
	\label{fig:keppel}
\end{figure}

\begin{figure}
\centering
    \includegraphics[width=0.5\textwidth]{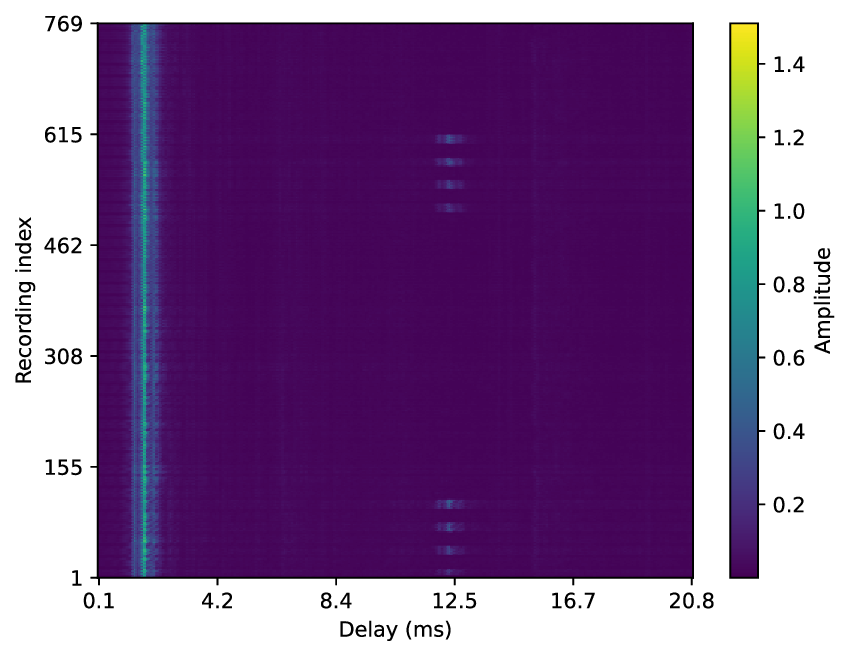}
    \caption{Amplitude profile of measured TVIRs in the complete Keppel dataset.}
    \label{fig:Keppel_amp}
\end{figure}

\begin{figure*}[t]
\centering
\includegraphics[width=\textwidth]{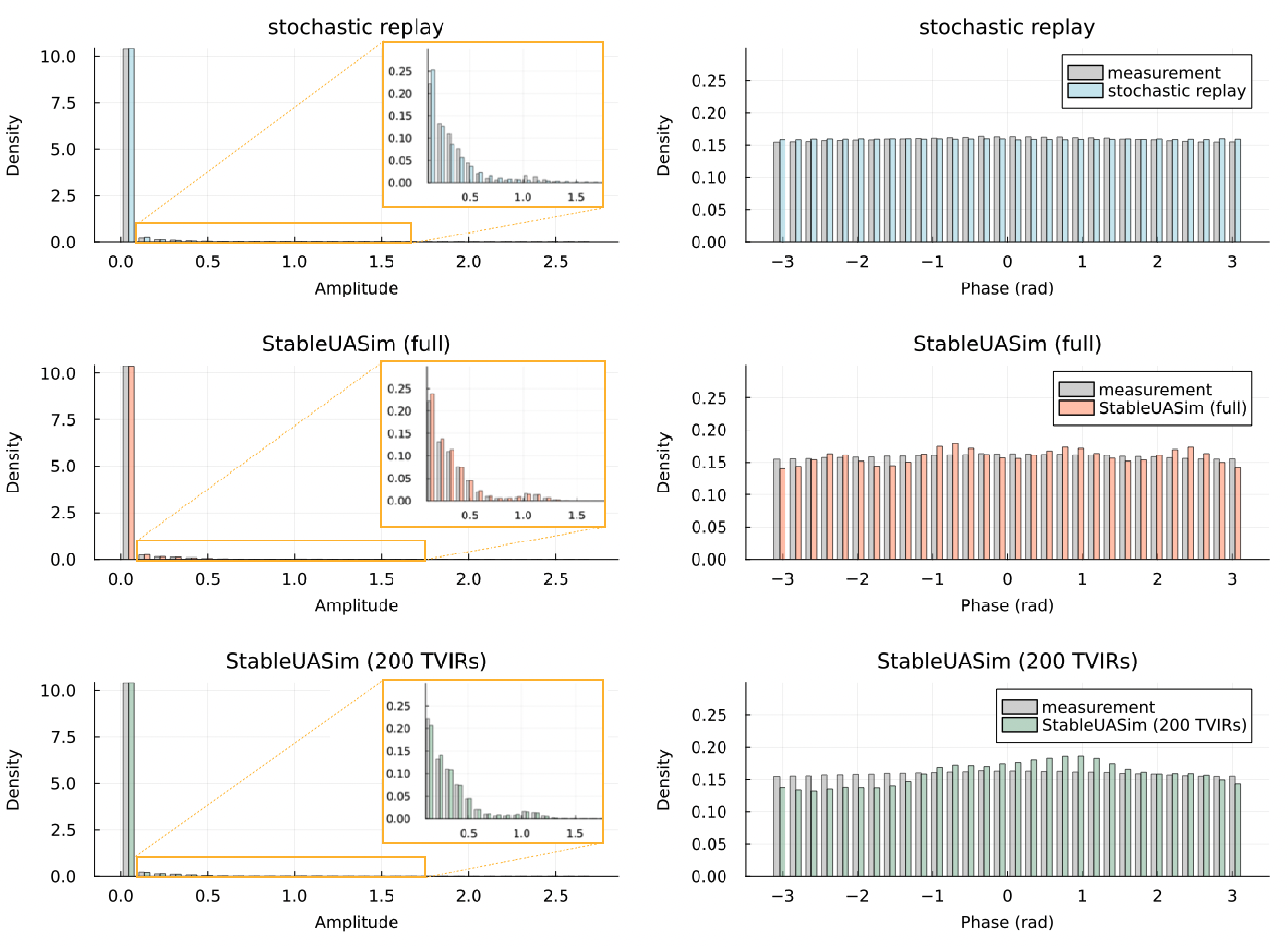}
\caption{\update{Histogram of significant tap amplitudes and phases for measured and generated channels. \emph{StableUASim~(full)} refers to the StableUASim model fine-tuned by full set of recordings. \emph{StableUASim~(\update{200 TVIRs})} denotes the model fine-tuned with 200 TVIRs (50 recordings $\times$ 4 channels/recording). The density is computed using normalized histograms, where counts are normalized by the total number of samples and bin width.}}
	\label{fig:keppel_hist}
\end{figure*}

A transmitter was deployed 150~m from a four-channel receiver array at a water depth of approximately 5~m, as shown in Fig.~\ref{fig:keppel}. In each cycle, the transmitter sent a probe signal of roughly 2~s duration, followed by three OFDM frames of 1~s each using a fixed set of OFDM parameters. The parameters were then adjusted, and the cycle repeated. In total, 192 recordings were recorded, resulting in 768 channel measurements across the four receiver channels.

Following a procedure analogous to the fine-tuning performed on the NOF1 dataset, all channels were used to further fine-tune the pre-trained baseline StableUASim model. The autoencoder was fine-tuned with an initial learning rate of 0.001, a patience of 10 epochs for reducing the learning rate by a factor of 10, and early stopping was triggered when the learning rate fell below $10^{-6}$. \update{For autoencoder fine-tuning, 90\% of the overall TVIRs are used for training, 20 TVIRs are reserved for testing, and the remaining 57 TVIRs are used for validation.} The diffusion model was fine-tuned with an initial learning rate of 0.005 and a patience of 50 epochs, with the learning rate reduced by a factor of 10 and training terminated once it fell below $10^{-6}$. \update{A fixed validation set consisting of 768 pairs, matching the size of the training dataset, was used and kept constant across epochs to provide a stable reference for monitoring and guiding early stopping.} Both models were optimized using the ADAM optimizer \update{and batch size of 128}. Fine-tuning of the diffusion model followed the full generative procedure described in Section~\ref{sec:NOF}.

We include a StableUASim variant fine-tuned with only 50 recordings across the 4 channels, yielding a total of 200 TVIRs, compared to the full set of 768 recorded TVIRs. \update{To avoid information leakage across channels when evaluating performance under limited data, the data split is performed at the recording level, ensuring that all four-channel TVIRs from each selected recording are assigned consistently to the training and validation set.} This variant is shown alongside the full-data StableUASim and baseline methods in all subsequent results, illustrating the model's ability to generalize from limited fine-tuning data. Unlike StableUASim, the direct replay method requires new channel measurements to increase diversity.

\begin{figure*}[]
    \includegraphics[width=\textwidth]{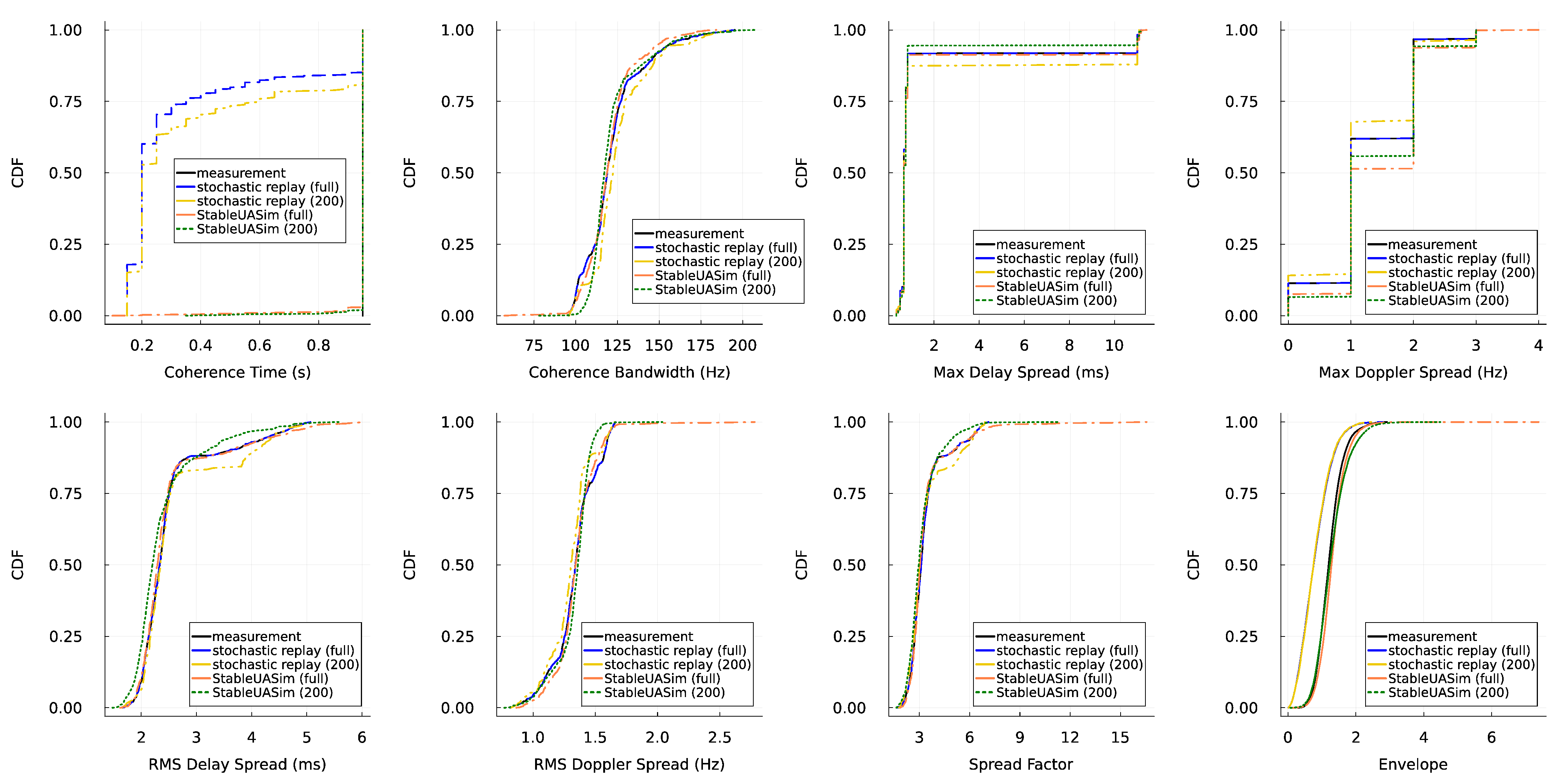}
\caption{CDF of eight key channel characteristics of Keppel Marina dataset. \update{Stochastic replay (200) and StableUASim (200) denote results fine-tuned using 200 TVIRs derived from the 50 recordings.}}
	\label{fig:keppel_cdf}
\end{figure*}

\begin{figure*}[]
    \includegraphics[width=\textwidth]{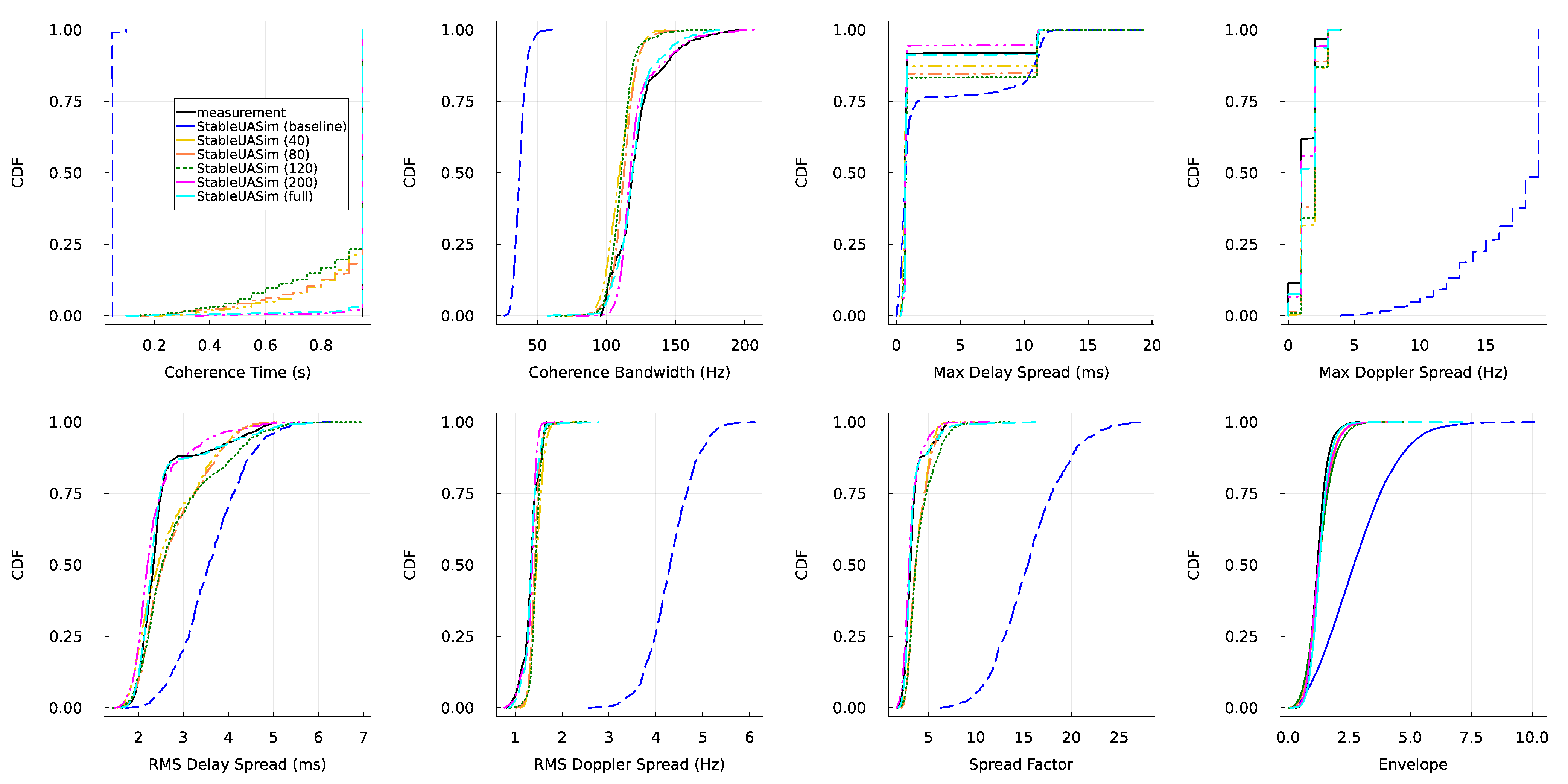}
\caption{\update{CDFs of eight key channel characteristics for the Keppel Marina dataset, comparing the baseline model (no fine-tuning) with models fine-tuned using 40, 80, 120, 200, and the full set of TVIRs.}}
	\label{fig:keppel_cdf_n_finetune}
\end{figure*}

To evaluate the generated channels, we first analyze the distributions of significant tap amplitudes and phases. As shown in Fig.~\ref{fig:keppel_hist}, the histograms of both StableUASim variants closely align with those of the measured channels, demonstrating strong agreement. In contrast, stochastic replay exhibits noticeable deviations. We then examine key channel statistical characteristics using CDFs, presented in Fig.~\ref{fig:keppel_cdf}. \new{The CDF comparison provides an evaluation of whether the generated channels preserve the statistical diversity of the full measurement dataset. The StableUASim-generated channels closely match the statistical characteristics of the measured dataset. This demonstrates that the proposed model is able to approximate the statistical diversity of the full dataset using only 200 TVIRs for fine-tuning. In contrast,} stochastic replay deviates most prominently in coherence time, where the CDF is shorter than that of the measurements. \new{This is because it relies on the observed samples and therefore cannot expand the diversity beyond the available measurements.} \update{To evaluate the effectiveness of fine-tuning, as shown in Fig.~\ref{fig:keppel_cdf_n_finetune}, we fine-tune the model using varying numbers of TVIRs and compare the resulting CDFs. The results show that even with a limited number of TVIRs, the model is able to capture the key channel characteristics, with performance further improving as more fine-tuning data is used.}

\FloatBarrier

\begin{figure*}[]
    \includegraphics[width=\textwidth]{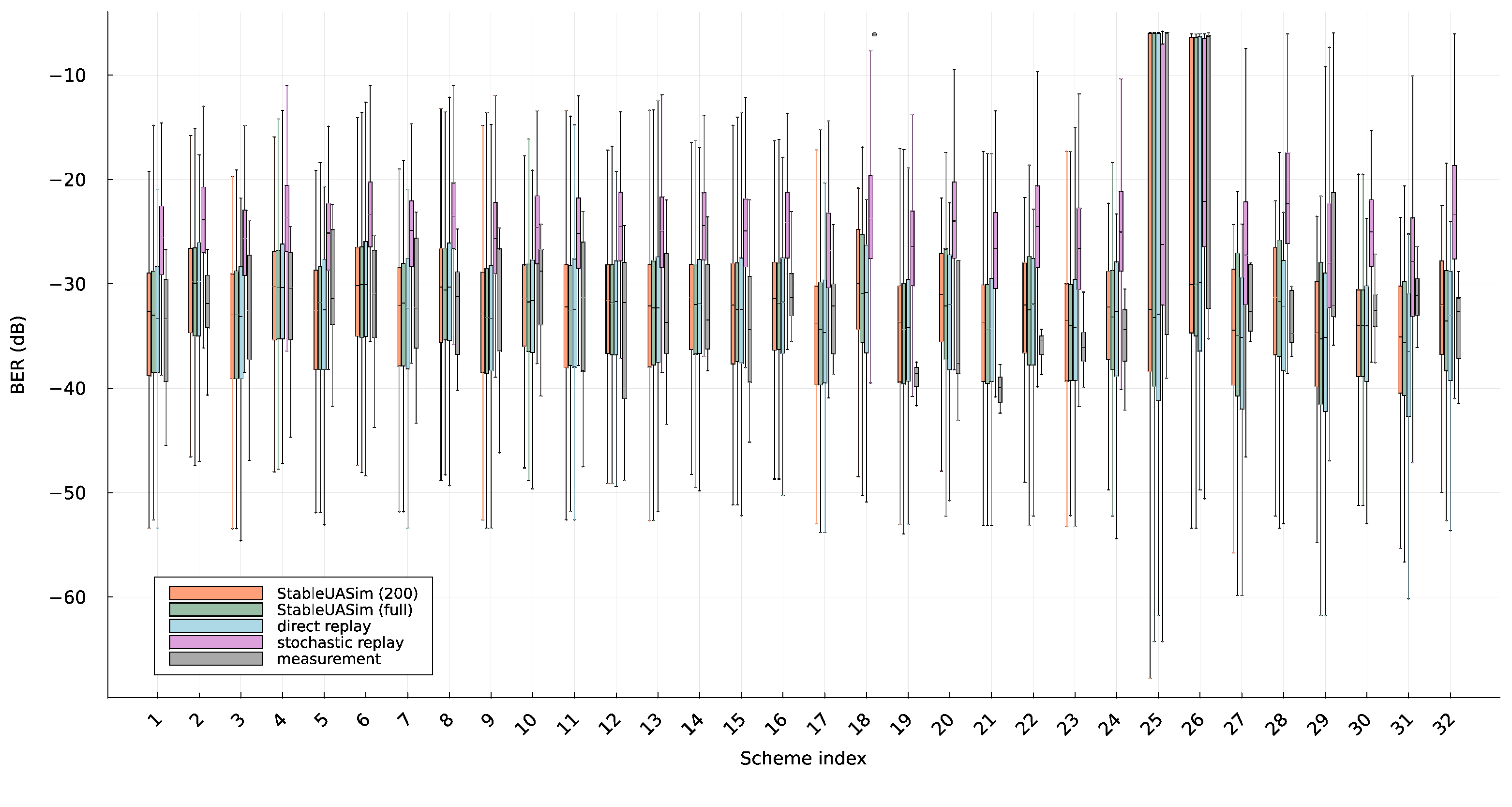}
\caption{\update{Box plot of BER (in dB, 75th percentile) across 32 communication schemes for measured, replayed, and generated channels. BER is evaluated using the same set of channel realizations across all schemes, with no scheme-specific retraining. Detailed parameters for each scheme are provided in Table~\ref{tab:scheme_code}.}}
	\label{fig:keppel_ber}
\end{figure*}

\begin{table*}
\centering
\small
\renewcommand{\arraystretch}{1.2}
\begin{tabular}{|l|l|l|l|l|l|l|l|}
\hline 
\textbf{Scheme} & \textbf{Frame} & \textbf{Bit per} & \textbf{Number of}& \textbf{Center carrier} & \textbf{Number} & \textbf{Number of cyclic } &  \\
\textbf{index} & \textbf{length} &\textbf{OFDM block} & \textbf{null carrier} & \textbf{number} & \textbf{of carrier} & \textbf{prefix length} & \textbf{Windowed} \\
\hline
1 & 586 & 102 & 307 & 21 & 512 & 0 & True \\
\hline
2 & 1173 & 204 & 307 & 21 & 512 & 0 & False \\
\hline
3 & 471 & 102 & 307 & 21 & 512 & 128 & True \\
\hline
4 & 943 & 204 & 307 & 21 & 512 & 128 & False \\
\hline
5 & 395 & 102 & 307 & 21 & 512 & 256 & True \\
\hline
6 & 790 & 204 & 307 & 21 & 512 & 256 & False \\
\hline
7 & 293 & 102 & 307 & 21 & 512 & 512 & True\\
\hline
8 & 586 & 204 & 307 & 21 & 512 & 512 & False \\
\hline
9 & 586 & 204 & 614 & 43 & 1024 & 0 & True\\
\hline
10 & 1173 & 408 & 614 & 43 & 1024 & 0 & False \\
\hline
11 & 525 & 204 & 614 & 43 & 1024 & 128 & True \\
\hline
12 & 1071 & 408 & 614 & 43 & 1024 & 128 & False\\
\hline
13 & 484 & 204 & 614 & 43 & 1024 & 256 & True\\
\hline
14 & 969 & 408 & 614 & 43 & 1024 & 256 & False\\
\hline
15 & 408 & 204 & 614 & 43 & 1024 & 512 & True \\
\hline
16 & 816 & 408 & 614 & 43 & 1024 & 512 & False\\
\hline
17 & 612 & 408 & 1228 & 85 & 2048 & 0 & True \\
\hline
18 & 1227 & 818 & 1228 & 85 & 2048 & 0 & False \\
\hline
19 & 561 & 408 & 1228 & 85 & 2048 & 128 &  True \\
\hline
20 & 1124 & 818 & 1228 & 85 & 2048 & 128 & False \\
\hline
21 & 510 & 408 & 1228 & 85 & 2048 & 256 &  True \\
\hline
22 & 1022 & 818 & 1228 & 85 & 2048 & 256 & False \\
\hline
23 & 459 & 408 & 1228 & 85 & 2048 & 512 &  True \\
\hline
24 & 920 & 818 & 1228 & 85 & 2048 & 512 &  False \\
\hline
25 & 613 & 818 & 2457 & 171  & 4096 & 0 & True \\
\hline
26 & 1227 & 1636 & 2457 & 171 & 4096 & 0 & False\\
\hline
27 & 613 & 81 & 2457 & 171 & 4096 & 128 &  True \\
\hline
28 & 1227 & 1636 & 2457 & 171 & 4096 & 128 &False\\
\hline
29 & 613 & 818 & 2457 & 171 & 4096 & 256 & True\\
\hline
30 & 1227 & 1636 & 2457 & 171 & 4096 & 256 & False \\
\hline
31 & 511 & 818 & 2457 & 171 & 4096 & 512 & True \\
\hline
32 & 1022 & 1636 & 2457 & 171 & 4096 & 512 &False \\
\hline
\end{tabular}
\caption{Detailed parameters of the 32 OFDM BPSK communication schemes evaluated on the Keppel Marina dataset. In the Windowed OFDM schemes, we apply a Hamming window instead of the rectangular pulse used in standard OFDM. By leaving alternate subcarriers unused (i.e., no energy transmitted), interference from adjacent carriers is reduced. This design improves robustness by making the system less sensitive to Doppler effects.}
\label{tab:scheme_code}
\end{table*}

Finally, we assess communication performance by estimating the BER of a given scheme over a channel represented by a TVIR using the real-time underwater acoustic simulator. The simulator encodes and transmits the signal using the selected scheme at a specified SNR, then demodulates the received signal to compute the BER for that channel and scheme. Fig.~\ref{fig:keppel_ber} presents a box plot of the BERs for both measured and generated channels. The direct replay method serves as the baseline BER, using the measured channel directly with additive Gaussian noise. Since the number of actual measurements for a particular scheme is very limited, the box plot uses all 768 TVIR channels across the four receivers, making it reasonable that the displayed measurements represent only a subset of the possible channel performance for each scheme. The StableUASim-generated channels accurately reproduce the measured channel properties, whereas stochastic replay deviates most prominently in coherence time, where the CDF is shorter than that of the measurements.

The discrepancies observed in stochastic replay stem from its underlying assumptions: (i) the UAC channel can be decomposed into a slow-varying trend (large-scale fading) and a fast-varying component (small-scale multipath fading), and (ii) once the slow trend is removed, the fast-varying component is assumed to be stationary with consistent statistics throughout the measurement period. These assumptions hold only if the trend and stationary components can be reliably separated and their statistics remain stable across the observed TVIRs. To achieve this decomposition, stochastic replay employs empirical mode decomposition (EMD)~\cite{huang1998empirical} to extract the slow- and fast-fading terms. However, for short-duration TVIRs, as in our Keppel Marina dataset, this process can be inaccurate, introducing artifacts that deviate from the true channel behaviour. This highlights a broader limitation of the stochastic replay method: its effectiveness strongly depends on the validity of the stationarity and decomposition assumptions.

\update{Across eight channel metrics, the empirical CDFs of the Keppel and NOF1 datasets exhibit broadly similar shapes, indicating shared underlying propagation characteristics. At the same time, consistent but subtle differences are observed: the Keppel dataset shows a slightly smaller coherence bandwidth range and larger RMS delay spread and Doppler spread, indicating more frequency-selective and time-varying channel characteristics. These differences primarily arise from environmental variability. Figs.~\ref{fig:NOF_cdf} and Fig.~\ref{fig:keppel_cdf} collectively illustrate the shared statistical structure as well as dataset-specific variations that the proposed model is required to generalize across.}

\section{Limitations and Future Work}
\label{sec:discussion}

Despite these promising results, several limitations remain. First, the autoencoder is trained on TVIRs with fixed sequence length and fixed delay taps. While it is sufficient for environments such as NOF1 and Singapore, this setup may struggle in channels with much longer multipath. Although LSTMs can handle variable-length sequences, this capability is not currently exploited. Future work could explore architectures and training strategies that support dynamic sequence lengths and delay spreads, potentially leveraging more diverse experimental pre-training datasets to enhance robustness and enable rapid adaptation to new environments.

Second, the training objectives are not directly aligned with the ultimate purpose of the surrogate channel model. The autoencoder minimizes reconstruction error, while the diffusion model minimizes noise prediction error in latent space. Although these objectives preserve signal fidelity, they do not explicitly optimize for communication-relevant metrics, such as delay spread, Doppler dynamics, or BER. Furthermore, not all reconstruction errors are equally consequential: some may be negligible, whereas others can significantly impair performance. Ideally, the model should prioritize features most relevant to end-to-end communication accuracy. 

\new{Third, this work focuses on sparse underwater acoustic channels, and the pre-training dataset is designed accordingly. Extending the pre-training dataset with more diverse simulated and experimental channel responses represents an important future direction to further improve the generalization capability of StableUASim across a broader range of underwater propagation environments.}

Looking forward, transformer architectures present a promising direction. With attention mechanisms, they can accommodate variable-length sequences and capture long-term temporal dependencies. An end-to-end transformer trained directly on task-relevant metrics could improve both adaptability across environments and accuracy of the surrogate model. More broadly, one can envision a large-scale generative \emph{underwater acoustic channel GPT} pre-trained on diverse simulated and experimental datasets. By incorporating structured environmental descriptors, such as water depth, seabed properties, range, and sound speed profile, such a model could flexibly generate realistic TVIRs for unseen conditions, analogous to how language models respond to prompts, bringing us closer to a general-purpose underwater acoustic channel simulator.

\section{Conclusion}

Surrogate modeling of UAC channels is both critical and challenging. Conventional physics-based propagation models require detailed environmental knowledge, while stochastic replay methods are constrained by restrictive assumptions and the limited availability of diverse measured instances, making them difficult to generalize to unseen environments. Generative machine learning models offer a promising alternative to overcome these challenges.

In this work, we introduced StableUASim, a conditional latent diffusion surrogate model tailored for UAC channels. By pre-training on diverse simulated environments, StableUASim achieves rapid adaptation to new conditions and generates high-fidelity TVIRs. Experimental validation on two real-world datasets demonstrates that, under the same available information, StableUASim achieve matches or even surpasses the state-of-the-art generative and stochastic replay models. More importantly, in scenarios where existing generative ML surrogate models and replay methods fail to function reliably, StableUASim remains robust and effective. 

While several challenges remain open, as discussed in Section~\ref{sec:discussion}, this study \new{demonstrates} how pre-trained generative models can benefit the underwater acoustic domain. Overall, StableUASim represents a significant step toward reliable, data-efficient, and scalable surrogate modeling, bridging the gap between physical modeling and standard ML for future underwater communication system design.

\bibliographystyle{IEEEtran}
\bibliography{references}

\end{document}